\documentclass[pra,twocolumn,tightenlines,nofootinbib,nolongbibliography]{revtex4-2}
\usepackage{bm,dcolumn,amsmath,graphicx,amsfonts,amssymb}

\usepackage{hyperref}
\usepackage[table,xcdraw]{xcolor}
\definecolor{cite}{rgb}{0.,0.,0.9}   
\hypersetup{colorlinks,linkcolor={cite},citecolor={cite},urlcolor={cite}}

\renewcommand{\v}[1]{\ensuremath{\boldsymbol{#1}}}

\newcommand{\bra}[1]{\ensuremath{\langle #1|}}
\newcommand{\ket}[1]{\ensuremath{|#1\rangle}}

\newcommand{\threej}[6]{
\renewcommand\arraystretch{0.75}\setlength\arraycolsep{1pt}
\small\ensuremath{\begin{pmatrix}#1&#2&#3\\#4&#5&#6\end{pmatrix}}}
\newcommand{\sixj}[6]{\renewcommand\arraystretch{0.75}
\setlength\arraycolsep{2pt}\small\ensuremath{
\begin{Bmatrix}#1&#2&#3\\#4&#5&#6\end{Bmatrix}}}

\def\d{\ensuremath{{\rm d}}}
\newcommand{\en}{\ensuremath{\varepsilon}}
\renewcommand{\a}{\ensuremath{\alpha}}
\newcommand{\w}{\ensuremath{\omega}}

\def\+{\ensuremath{{^+}}}

\newcommand{\smallspace}{\rule{0pt}{2.5ex}}

\usepackage[normalem]{ulem}
\definecolor{newc}{rgb}{0.,0.6,0.4}

\usepackage{adjustbox}

\usepackage{lipsum}

\begin{document}

\title{Electric dipole transition amplitudes for atoms and ions with one valence electron}

\author{B.\ M.\ Roberts}\email[]{b.roberts@uq.edu.au}
\author{C.\ J.\ Fairhall}
\author{J.\ S.\ M.\ Ginges}
\affiliation{School of Mathematics and Physics, The University of Queensland, Brisbane QLD 4072, Australia}
\date{\today}

\begin{abstract}\noindent
Motivated by recent measurements for several alkali-metal atoms and alkali-metal-like ions, we perform a detailed study of electric dipole (E1) transition amplitudes in K, Ca\+, Rb, Sr\+, Cs, Ba\+, Fr, and Ra\+, which are of interest for studies of atomic parity violation, electric dipole moments, and polarizabilities. 
Using the all-orders correlation potential method, we perform high-precision calculations of E1 transition amplitudes between low-lying $s$, $p$, and $d$ states. 
We perform a robust error analysis, and compare our calculations to many amplitudes for which there are high-precision experimental determinations.
We find excellent agreement, with deviations at the level of $\sim$\,0.1\%.
We also compare our results to other theoretical evaluations, and discuss the implications for uncertainty analyses.
Further, combining calculations of branching ratios with recent measurements, we extract high-precision values for several E1 amplitudes of Ca\+, Sr\+, Cs, Fr, and Ra\+. 
\end{abstract}

\maketitle

\section{Introduction}

\noindent
Recent advances in experimental and theoretical techniques have created a new era of unprecedented precision in the study of atomic phenomena.
Atomic physics plays an ever-growing role in fundamental physics studies,
including through atomic parity violation and searches for permanent electric dipole moments~\cite{GingesRev2004,RobertsReview2015}, as well as for tests of the CPT theorem and Lorentz symmetry, searches for variation of fundamental constants, and detection of dark matter and dark energy~\cite{AtomicReview2017}.

This paper is partly motivated by recent high-precision measurements of alkali-like atoms and ions,
including lifetimes for Cs~\cite{Toh2018,Pucher2020} and Ra\+~\cite{Fan2022}, and new spectroscopic measurements for 
Rb~\cite{Fallon2022},
Cs~\cite{Damitz2019,Toh2019a,Toh2019,TohBeta2019,Quirk2022,*Quirk2022c},
Ca\+~\cite{Meir2020,Song2019},
Ba\+~\cite{Dijck2018,Arnold2020,Zhang2020a} and 
Ra\+~\cite{Fan2019,Fan2019a,Holliman2020}.
There are recent proposals and experimental advances towards developing optical atomic clocks in, e.g.,
Ca\+~\cite{Zhang2020b,*Zhang2020c}, 
Rb~\cite{Gerginov2018,Perrella2019,Martin2019,HamiltonRb2022}, %
Sr\+~\cite{Barwood2014,*Dube2015}, 
Cs~\cite{Sharma2022},
and 
Ra\+~\cite{Holliman2022,Holliman2022a}.
There has been recent progress towards a new measurement of atomic parity violation (APV) in Cs~\cite{TohBeta2019,Katsoprinakis2019,Choi2018,Choi2016,Toh2014,Antypas2014,*
Antypas2013a,*Antypas2013b,Lintz2007,*Guena2005},
and there is ongoing work towards APV experiments in Fr~\cite{Gwinner2021,Bouchiat2008,Tandecki2013,Aubin2013}, Ra\+~\cite{Fan2019,NunezPortela2014,*NunezPortela2013,Versolato2011,Traykov2008}, and Ba\+~\cite{Koerber2003,Fortson1993}.
It is important that new APV measurements be performed to check and improve upon the current most-precise Cs measurement~\cite{Wieman1997}.

There have also been improvements in theoretical methods, including the complete inclusion of triple excitations into coupled-cluster calculations~\cite{Porsev2006,Derevianko2008,Porsev2010,Porsev2021b,Porsev2021,Sahoo2021}, and new techniques for computing APV amplitudes~\cite{Tan2021b}.
Further, our recent study of quantum electrodynamics (QED) corrections to electric dipole (E1) amplitudes demonstrated that in several cases the QED correction is larger than the discrepancy between theory and experiment~\cite{FairhallQED2022}.

Calculations of E1 matrix elements are required to interpret APV measurements, both through a direct calculation of the APV amplitude~\cite{DzubaCPM1989plaPNC,Blundell1992,GingesCs2002,ShabaevPRL2005,*Shabaev2005,
Porsev2009,OurCsPNC2012,Sahoo2021,RobertsComment2021,Tan2021b}, and for the determination of transition polarizabilities required to extract the nuclear weak charge~\cite{Bouchiat1974,*Bouchiat1975}. 
Currently, atomic theory  is the limiting factor in studies of APV~\cite{RobertsReview2015}. 
Further, there is a $2.8\,\sigma$ tension in the Cs $6s$-$7s$ vector transition polarizability value derived via two methods~\cite{TohBeta2019,Cho1997,DzubaCs2000,Wieman1999},
both of which require atomic theory input.

Accurate E1 amplitudes are also important in many other areas, including searches for new physics in other precision experiments~\cite{AtomicReview2017}, 
astrophysical analyses~\cite{Pehlivan2015,Gallagher2020}, and 
studies of polarizabilities~\cite{Mitroy2010}
(which in turn are important for
the development of atomic clocks~\cite{DereviankoRMP2011,*Ludlow2015}, 
studies of long-range atomic interactions~\cite{Porsev2014}, and quantum information~\cite{Goldschmidt2015}). 
Moreover, E1 amplitudes provide a benchmark for testing atomic theory.
Comparisons between theory and experiment probe the accuracy of wavefunctions at large distances from the nucleus, complementary to 
hyperfine comparisons, which probe the accuracy at small distances~\cite{Ginges2018,RobertsHFA2021,SanamyanMuon2022}.

In this work, we perform high-precision calculations of E1 amplitudes between $s$, $p$, and $d$ states of 
the alkali-metal atoms 
potassium, rubidium, cesium, and francium, and the alkali-like singly-ionized calcium, strontium, barium, and radium.
We compare our results to 46 amplitudes in these systems for which there are high-precision experimental data available, finding excellent agreement. 
The typical differences are at the level of $\sim$\,0.1\%;
more than half of our calculated amplitudes {\sl lie within experimental uncertainties}. 
This sets a new precedent for state-of-the-art atomic theory.
We use a robust method to determine the theory uncertainties, and demonstrate statistically that these are conservative:\ all but two amplitudes lie within combined theory and experimental uncertainties, better than statistically expected.
We also compare our results to other theoretical evaluations; there is good agreement among most of the highest-precision calculations with a few exceptions that will be discussed below.

\section{Theory}

We employ a new implementation of the all-orders correlation potential method first introduced in Refs.~\cite{DzubaCPM1988pla,DzubaCPM1989plaEn}.
We refer to our implementation as atomic many-body perturbation theory in the screened Coulomb interaction (\textsc{ampsci}).
We outline the main aspects of the method here; details are presented in the appendix.
Except where noted, we use atomic units
($\hbar$\,=\,$e$\,=\,$m_e$\,=\,$c\alpha$\,=\,$1$)
with the Gaussian system of electromagnetic units. 

\subsection{Correlation corrections}

We consider $N$-electron atoms and ions that have a single valence-electron above a closed noble-gas--like core.
A natural starting point is the relativistic Hartree-Fock (RHF) model, with the wavefunction for the valence electron found in the frozen potential of the $N$\,$-$\,1 core electrons.
A non-local energy-dependent operator, $\Sigma$, is added to the RHF equation for the valence state to account for the core-valence correlations~\cite{DzubaPNC1984,*DzubaPNC1985}:
\begin{equation}\label{eq:H-Bru}
(h_{\rm HF} +  \Sigma)\psi^{\rm (Br)} = \en^{\rm (Br)}\psi^{\rm (Br)},
\end{equation}
where $h_{\rm HF}$ is the single-particle RHF Hamiltonian.
The resulting orbitals are known as Brueckner orbitals.

\begin{figure}
\centering\tiny
\includegraphics[width=0.11\textwidth]{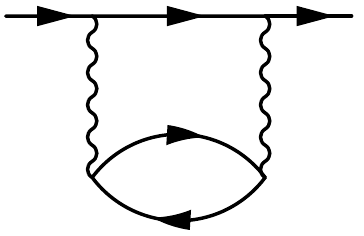}
\includegraphics[width=0.11\textwidth]{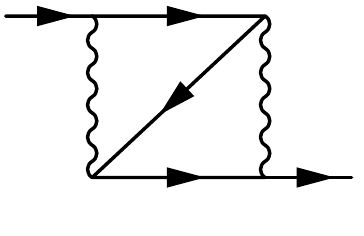}
\includegraphics[width=0.11\textwidth]{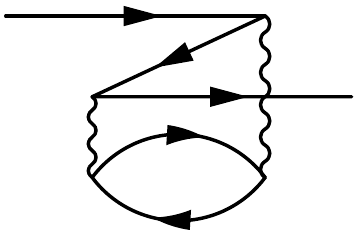}
\includegraphics[width=0.11\textwidth]{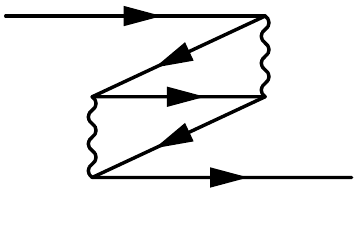}
\caption{\small Goldstone diagrams for the second-order energy correction. Backward facing lines denote states in the core (holes) and internal forward lines are virtual excited states. 
\label{fig:Sigma2}}
\end{figure}

\begin{figure}
\centering
\includegraphics[height=0.025\textwidth]{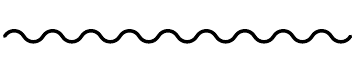}
\raisebox{0.012\linewidth}{\large$+$}
\includegraphics[height=0.025\textwidth]{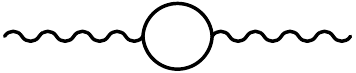}
\raisebox{0.012\linewidth}{\large$+$}
\includegraphics[height=0.025\textwidth]{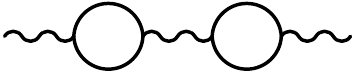}
\raisebox{0.012\linewidth}{\large$+$}
\raisebox{0.012\linewidth}{\large$\ldots$}
\caption{\label{fig:Screening}\small Screening of the Coulomb interaction by polarization of the core (using the Feynman technique).}
\end{figure}

\begin{figure}
\centering
\includegraphics[height=0.04\textwidth]{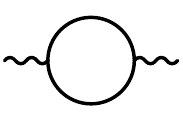}
\raisebox{0.03\linewidth}{\large$+$}
\includegraphics[height=0.04\textwidth]{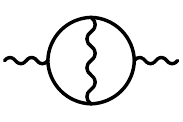}
\raisebox{0.03\linewidth}{\large$+$}
\includegraphics[height=0.04\textwidth]{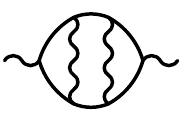}
\raisebox{0.03\linewidth}{\large$+$}
\raisebox{0.02\linewidth}{\large$\ldots$}
\caption{\label{fig:HoleParticle}\small 
Hole-particle corrections to the polarization operator.}
\end{figure}

\begin{figure}
\centering
\includegraphics[height=0.0285\textwidth]{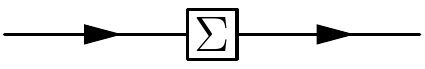}~
\raisebox{0.015\linewidth}{\large$+$}
\includegraphics[height=0.0285\textwidth]{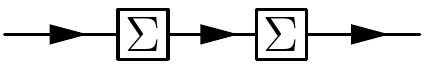}~
\raisebox{0.015\linewidth}{\large$+~\ldots$}
\caption{\label{fig:Chaining}\small 
Example $\Sigma$ chaining diagrams.}
\end{figure}

It is possible to calculate $\Sigma$ to lowest (second) order in the Coulomb interaction as in Fig.~\ref{fig:Sigma2} (see, e.g., Ref.~\cite{Lindgren1986}),
which we refer to as $\Sigma^{(2)}$.
For accurate calculations, correlations beyond second-order are required.
We use the Feynman diagram technique developed in Refs.~\cite{DzubaCPM1988pla,DzubaCPM1989plaEn},
in which the dominating correlations (screening of the Coulomb interaction by the core electrons, shown in Fig.~\ref{fig:Screening}, and the hole-particle interaction, shown in Fig.~\ref{fig:HoleParticle}) are included to all orders.
By solving the Brueckner equation \eqref{eq:H-Bru}, a third set of diagrams (chaining of the $\Sigma$ operator, shown in Fig.~\ref{fig:Chaining}) is also included to all orders~\cite{DzubaPNC1984,*DzubaPNC1985}.
We refer to the all-orders correlation potential as $\Sigma^{(\infty)}$.

\subsection{Interaction with external fields}

In the presence of an external field oscillating with frequency $\omega$, the atomic orbitals contain perturbations,
\begin{equation}\label{eq:tdhf-dPsi}
\psi \to \psi + \delta\psi = \psi +  X e^{- i\w t}+ Y e^{ i\w t}
\end{equation}
and $\en\to\en+\delta\en (e^{- i\w t}+ e^{ i\w t})$.
This leads to corrections to matrix elements known as core polarization.
To first order in the external field, the corrections satisfy the time-dependent Hartree-Fock (TDHF) equations~\cite{DzubaHFS1984}
\begin{equation}\begin{split}
\label{eq:tdhf}
\left( h_{\rm HF} - \en -\w \right)X &= -\left( t^k_q + \delta V - \delta \en \right)\psi \\
\left( h_{\rm HF} - \en +\w \right)Y &= -\left( {t^{k\dag}_q} + \delta V^\dag - \delta \en \right)\psi ,
\end{split}\end{equation}
with $\delta\en = \bra{\psi} t^k_q + \delta V\ket{\psi}$,
where $t^k_q$ is the operator of the field (with rank $k$ and projection $q$),
$X$ and $Y$ are generally not states of definite angular momentum, 
and $\delta V$ is the resulting correction to the RHF potential:
\begin{equation}\label{eq:delV-simple}
\delta V = V_{\rm HF}(\{\psi_c + \delta\psi_c\}) - V_{\rm HF}(\{\psi_c\}).
\end{equation}

The set of THDF equations (\ref{eq:tdhf}) and (\ref{eq:delV-simple}) are solved self-consistently for the core; thereby, core polarization is included to all orders in the Coulomb interaction.
Matrix elements including core polarization are then calculated as
$
\bra{w} t^k_q + \delta V\ket{v}
$~\cite{DzubaHFS1984}.
The TDHF method is equivalent to the diagrammatic random phase approximation with exchange (RPA), see, e.g., Ref.~\cite{Johnson1980}, though the TDHF method is typically more numerically stable, does not require a basis, and automatically includes contributions from negative energy states.
In the present case, the external field is that driving the E1 transition, $\omega$ is the transition frequency, 
$t^k$\,$=$\,$\v{d}$\,$=$\,$-e\v{r}$ ($k$\,=\,1), 
and $\delta\en=0$. %

\subsection{Non-Brueckner correlations}

\begin{figure}
\centering
\includegraphics[width=0.155\textwidth]{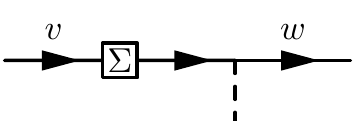}
\includegraphics[width=0.155\textwidth]{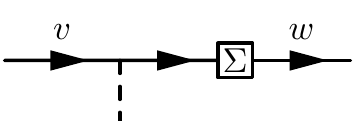}
\includegraphics[width=0.155\textwidth]{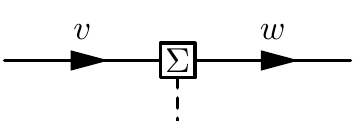}
\caption{\label{fig:SR1}\small Example correlation corrections to matrix elements:\ $\Sigma$ is the correlation potential, and the dashed line is 
an external field. 
The first two diagrams are included via the use of Brueckner orbitals for the valence states;
 the third diagram, structure radiation, is calculated separately.}
\end{figure}

By calculating matrix elements as described above, the most important correlation and core-polarization effects are included.
However, there are other many-body corrections that cannot be incorporated in that way.
The most important of these are structure radiation (SR), which arises from the perturbation of the correlation potential by the external field as shown in Fig.~\ref{fig:SR1}, and normalization of states (NS), which arises due to the change of the normalization of the many-body wavefunction~\cite{Dzuba1987jpbRPA,Johnson1987}.
These corrections enter at roughly the same level ($\lesssim1\%$) and tend to cancel.
Nevertheless, for high-precision calculations they must be included.

Expressions for SR were presented, e.g., in Ref.~\cite{JohLiuSap96}:
\begin{align}\label{eq:SR}
\delta t^{\rm SR}_{wv}
= & \sum_{ar} t_{ar}\left[\frac{ T^k_{wrva}}{\en_r - \en_a + \omega}
+ \frac{   (-1)^{j_v-j_w}T^k_{vrwa}}{\en_r - \en_a - \omega}\right]\notag\\
 &-\sum_{ab} {t_{ba} C^{k}_{wavb}} - \sum_{rn} {t_{rn} D^{k}_{wnvr}},
\end{align}
where 
$t_{ij}$\,$\equiv$\,$\bra{i}|t^k|\ket{j}$,
{${a,\,b,...}$} are core states, $n,\,m,...$ are excited states, and {${v,\,w}$} are valence states.
Full expressions for 
$T^k$, $C^k$, and $D^k$ are given in Appendix~\ref{sec:A-SR+N}. 
The NS contribution may be expressed as~\cite{Dzuba1987jpbRPA}
\begin{equation}
\delta t^{\rm Norm}_{wv}
= \frac{-t_{wv}}{2}
\left[\bra{w}{\partial \Sigma}/{\partial \epsilon}\ket{w}  +  \bra{v}{\partial \Sigma}/{\partial \epsilon}\ket{v}\right].
\end{equation} 
In the lowest (third) order, the derivatives are determined analytically from the diagrams in Fig.~\ref{fig:Sigma2}.
The sums over intermediate states are performed using the dual-kinetic-balance B-spline basis as introduced in Ref.~\cite{Beloy2008}, which offers a high level of convergence and stability (see also Refs.~\cite{Johnson1988,ShaTupYer04}). 
We used 40 splines of order 7 in a cavity of $40\,a_0$ for the neutral atoms and $30\,a_0$ for the ions.
We have checked that basis truncation errors are negligible at the level of the claimed uncertainties.

\subsection{Radiative QED corrections}

For high-precision calculations it is important to include radiative QED corrections due to vacuum polarization and electron self-energy.
As demonstrated in Ref.~\cite{FairhallQED2022}, the QED effects can be pivotal for accurate calculations of E1 matrix elements in alkali atoms.

These may be incorporated using the radiative potential method~\cite{FlambaumQED2005}, in which an effective potential is added to the Hamiltonian. 
This is written as the sum of the Uehling (vacuum polarization),  high- and low-frequency electric self-energy, and magnetic self-energy potentials:
\begin{equation}\label{eq:Vrad}
V_{\rm rad}(\v{r}) = V_{\rm Ueh}(r) + V_{\rm h}(r) +  V_{\rm l}(r) +  i (\v{\gamma}\cdot\v{n}) V_{\rm m}(r).
\end{equation}
Full expressions, including finite-nuclear-size effects, are given in Ref.~\cite{GingesQED2015,*Ginges2016}.
An approximate form of the Wichmann-Kroll potential (higher-order vacuum polarization~\cite{Wichmann1956}) may also be added as in Ref.~\cite{FlambaumQED2005}, though is entirely negligible. 
Including $V_{\rm rad}$ into the RHF equations gives an important correction known as core relaxation, which dominates for states with $l>0$~\cite{Derevianko2004,GingesQED2015,*Ginges2016,FairhallQED2022}.

Using this method, QED radiative corrections can be included into atomic wavefunctions, and thereby into matrix elements~\cite{FlambaumQED2005,RobertsQED2013,FairhallQED2022}.
There are also QED contributions arising from corrections to the external field operator, e.g., vertex corrections
(similar to Fig.~\ref{fig:SR1}, with $\Sigma$ replaced by the radiative potential), which we do not include.
The self-energy vertex corrections for E1 amplitudes were estimated to be small in Ref.~\cite{FlambaumQED2005}; this was confirmed in Ref.~\cite{FairhallQED2022} by comparing with rigorous QED~\cite{Sapirstein2005} in simple atromic potentials.
While the QED corrections to E1 matrix elements may be calculated accurately using the radiative potential method, this is not the case in general (e.g., for hyperfine structure or APV matrix elements), see discussion in Refs.~\cite{RobertsQED2013,RobertsComment2021}.

\begin{table}
    \caption{Breit corrections to E1 reduced matrix elements and ionization energies for the lowest states of Cs as calculated in this work, and comparison with Porsev {\sl et al.}~\cite{Porsev2010}.
    \label{tab:Breit}}
    \begin{ruledtabular}
        \begin{tabular}{lD{.}{.}{2.4}D{.}{.}{2.4}D{.}{.}{2.4} c D{.}{.}{1.1}D{.}{.}{1.1}D{.}{.}{2.1}D{.}{.}{2.1}}
& \multicolumn{3}{c}{E1 ($ea_0)$}&& \multicolumn{4}{c}{Energy (cm$^{-1}$)}\\
\cline{2-4}\cline{6-9}
 & \multicolumn{1}{c}{$6s$-$6p_{1/2}$} &  \multicolumn{1}{c}{$6s$-$7p_{1/2}$} &  \multicolumn{1}{c}{$7s$-$6p_{1/2}$}
&& \multicolumn{1}{c}{${6s}$}& \multicolumn{1}{c}{${7s_{1/2}}$}& \multicolumn{1}{c}{${6p_{1/2}}$}& \multicolumn{1}{c}{${7p_{1/2}}$}
\\
\hline
\textsc{ampsci} & -0.0008 & 0.0017& 0.0046&&2.8&0.0&-7.4&-2.3\\
Ref.~\cite{Porsev2010}      & -0.0010 & 0.0019 & 0.0049&&2.6&0.3&-7.1&-2.5\\
        \end{tabular}
    \end{ruledtabular}
\end{table}

\subsection{Breit interaction}

The lowest-order relativistic correction to the electron-electron Coulomb interaction can be described by the the Breit Hamiltonian (see, e.g.,~Ref.~\cite{BetheBook}).
It enters as a correction to the electron Coulomb term in the many-body Hamiltonian:
${r_{ij}^{-1}}
\to
 r_{ij}^{-1} +  h^B_{ij}$,
where $r_{ij}$\,$=$\,$|\v{r}_i$$-$$\v{r}_j|$. 
In the limit of zero frequency, it may be expressed as
\begin{equation}
 h^B_{ij} = - \frac{\v{\a}_i\cdot\v{\a}_j + (\v{\a}_i\cdot\v{n}_{ij})(\v{\a}_j\cdot\v{n}_{ij})}{2\, {{r}}_{ij}},
\end{equation}
with $\v{n}_{ij}$\,=\,$(\v{r}_i$\,$-$\,$\v{r}_j)/r_{ij}$.
(Frequency-dependent effects can be neglected in most situations, however, they may become important for highly-charged ions~\cite{Johnson1988a,*Mann1971}.)
This leads to a correction to the RHF potential:\
$V_{\rm HF}$\,$\to$\,$V_{\rm HF}$\,+\,$V_{\rm B}$.
Including this into the RHF and TDHF equations allows for the inclusion of Breit effects into the calculations of atomic energies and wavefunctions~\cite{Derevianko2000,*Derevianko2001}.

As demonstrated in Refs.~\cite{Derevianko2001,DzubaCs2001}, including Breit in conjunction with correlations is very important. 
Our Breit corrections to the Cs energies and E1 matrix elements agree very well with those presented in Refs.~\cite{Derevianko2001,Porsev2010}, as shown in Table~\ref{tab:Breit}.
Breit corrections to E1 amplitudes for the Ba\+ ion can also be deduced from the calculations in Ref.~\cite{Porsev2021}; these agree precisely with ours. 

\subsection{Contribution from higher-order correlations}

\begin{table}
\caption{Ab initio calculations of ionization energies (cm$^{-1}$) for the lowest states of Cs, showing the breakdown of contributions;
$\Delta$ shows the deviation from experiment~\cite{NIST}.\label{tab:CsEn}}
\begin{ruledtabular}
\begin{tabular}{lrrrrrrrr}
 Level       & RHF      & $\delta\Sigma^{(2)}$     & $\delta\Sigma^{(\infty)}$   & Breit & QED   & {Final}   & Expt. & $\Delta$(\%)                      \\
\hline
$6s_{1/2}$ & 27954 & 4458 & -998 & 2.8  & -21.5 & 31395 & 31406 & -0.04\% \\
$6p_{1/2}$ & 18791 & 1747 & -294 & -7.4 & 1.1   & 20236 & 20228 & 0.04\%  \\
$6p_{3/2}$ & 18389 & 1550 & -258 & -0.7 & 0.1   & 19680 & 19674 & 0.03\%  \\
$5d_{3/2}$ & 14138 & 3424 & -458 & 25.8 & 5.6   & 17136 & 16907 & 1.4\%   \\
$5d_{5/2}$ & 14163 & 3240 & -402 & 30.3 & 4.7   & 17035 & 16810 & 1.3\%  \\
$7s_{1/2}$ & 12112 & 957  & -208 & 0.0  & -5.0  & 12856 & 12871 & -0.1\%  \\
$7p_{1/2}$ & 9223  & 506  & -81  & -2.6 & 0.4   & 9645  & 9641  & 0.04\%  \\
$7p_{3/2}$ & 9079  & 458  & -74  & -0.4 & 0.0   & 9463  & 9460  & 0.03\%  \\
\end{tabular}
\end{ruledtabular}
\end{table}

In Table~\ref{tab:CsEn}, we present {\sl ab inito} calculations of ionization energies for the lowest states of Cs.
This shows excellent agreement with experiment, particularly for $s$ and $p$ states.
The relatively worse agreement for the $d$ states is discussed below.
A similar level of agreement is found for the other systems. 

Finally, we  seek to estimate the impact of missed correlation effects. 
This is achieved by introducing semi-empirical scaling factors into the correlation potential in Eq.~(\ref{eq:H-Bru}),
$\Sigma\to\lambda\Sigma$,
which are tuned to reproduce the experimental energies.
The factors are extremely close to 1, due to the excellent accuracy of the {\sl ab initio} results. 
For example, the factors for the Cs $6s$ and $6p_{1/2}$ states are
$\lambda_{6s}\simeq 1.003$ and $\lambda_{6p_{1/2}}\simeq 0.995$.
They differ more for $d$ states, though are still very close to 1, $\lambda_{5d}\simeq 0.94$.
The effect of the scaling on matrix elements is small, though does lead to an improvement in the accuracy, see Table~\ref{tab:E1-lambda}.
We use this shift to gauge one of the main sources of uncertainty in the calculations 
as described below.
Importantly, the assigned uncertainty is always larger than the semi-empirical correction.

\begin{table}
    \caption{Example comparison of lowest Cs E1 reduced matrix elements across various approximations (RPA, Breit, QED, SR and NS contributions are included) Units: $ea_0$. \label{tab:E1-lambda}}
\begin{ruledtabular}
\begin{tabular}{lrrrrl}
\multicolumn{1}{c}{$|E1|$}
     & \multicolumn{1}{c}{$\Sigma^{(2)}$} & \multicolumn{1}{c}{$\lambda\Sigma^{(2)}$} &  \multicolumn{1}{c}{$\Sigma^{(\infty)}$} &  \multicolumn{1}{c}{$\lambda\Sigma^{(\infty)}$} &Expt.~\cite{TohBeta2019,Damitz2019}\\
\hline
$|\bra{6s}|d|\ket{6p_{1/2}}|$ &4.383 &4.496 & 4.507 & 4.505 & 4.5057(16) \\
$|\bra{6s}|d|\ket{6p_{3/2}}|$ &6.164 &6.327 & 6.343 & 6.340 & 6.3398(22) \\
$|\bra{6s}|d|\ket{7p_{1/2}}|$ &0.304 &0.282 & 0.273 & 0.278 & 0.27810(45) \\
$|\bra{6s}|d|\ket{7p_{3/2}}|$ &0.610 &0.580 & 0.570 & 0.574 & 0.57417(57) \\
\end{tabular}
\end{ruledtabular}
\end{table}

\begin{table*}
\caption{Calculations of reduced matrix elements (absolute values) between the lowest few states of Cs, and comparison with experiment (units:\ $ea_0$).
Here, $\delta V$ is the core polarization (RPA) correction, $\Sigma$ is the all-orders (Brueckner) correlation correction, SR+N is the combined structure radiation and normalization correction, $\delta\lambda$ is the scaling correction,
and $\Delta$ is the deviation from experiment ($\dagger$ means theory value lies within experimental uncertainty).
\label{tab:E1-Cs}}
\begin{ruledtabular}
\begin{tabular}{lrrrrrrrllD{.}{.}{1.4}}
         \multicolumn{1}{c}{$|\bra{-}|d|\ket{+}|$}
        &  \multicolumn{1}{c}{RHF}
        &  \multicolumn{1}{c}{$\delta V$}
        &  \multicolumn{1}{c}{$\delta\Sigma^{(\infty)}$}
        &  \multicolumn{1}{c}{Breit}
        & \multicolumn{1}{c}{QED}
        & \multicolumn{1}{c}{SR+N}
        & \multicolumn{1}{c}{$\delta\lambda$}
        &  \multicolumn{1}{c}{Final}
        &  \multicolumn{1}{c}{Expt.}
        &  \multicolumn{1}{c}{$\Delta$(\%)}                      \\
\hline\smallspace
 $6p_{1/2}$--$6s$ & 5.2777 & -0.3030 & -0.4634 & -0.0008 & 0.0034 & -0.0073 & -0.0015 & 4.5052(54) & 4.5057(16)~\cite{TohBeta2019}\tablenotemark[1] & -0.01\%^\dagger \\
 $6p_{3/2}$--$6s$ & 7.4264 & -0.4128 & -0.6641 & -0.0009 & 0.0051 & -0.0112 & -0.0023 & 6.3402(79) & 6.3398(22)~\cite{TohBeta2019}\tablenotemark[1] & 0.01\%^\dagger \\
 $7p_{1/2}$--$6s$ & 0.3717 & -0.1335 & 0.0170 & 0.0017 & -0.0023 & 0.0188 & 0.0043 & 0.2776(75) & 0.27810(45)~\cite{Damitz2019} & -0.2\% \\
 $7p_{3/2}$--$6s$ & 0.6947 & -0.1864 & 0.0396 & 0.0002 & -0.0026 & 0.0243 & 0.0044 & 0.5741(89) & 0.57417(57)~\cite{Damitz2019} & -0.01\%^\dagger \\
 $6p_{1/2}$--$7s$ & 4.4131 & 0.0368 & -0.1946 & 0.0046 & -0.0044 & -0.0313 & 0.0147 & 4.239(18) & 4.249(4)~\cite{Toh2019} & -0.2\% \\
 $6p_{3/2}$--$7s$ & 6.6710 & 0.0420 & -0.2112 & 0.0012 & -0.0054 & -0.0427 & 0.0191 & 6.474(23) & 6.489(5)~\cite{Toh2019} & -0.2\% \\
 $7p_{1/2}$--$7s$ & 11.0089 & -0.0877 & -0.5904 & -0.0023 & 0.0070 & -0.0166 & -0.0223 & 10.297(23) & 10.308(15)\tablenotemark[2] & -0.1\%^\dagger \\
 &&&&&&&&& 10.325(5)\tablenotemark[3] & -0.3\%\\
 $7p_{3/2}$--$7s$ & 15.3448 & -0.1172 & -0.8790 & -0.0005 & 0.0101 & -0.0227 & -0.0320 & 14.303(33) & 14.320(20)\tablenotemark[2] & -0.1\%^\dagger \\
 &&&&&&&&& 14.344(7)\tablenotemark[3] & -0.3\%\\
 $6p_{1/2}$--$5d_{3/2}$ & 8.9783 & -0.3397 & -1.6856 & -0.0107 & -0.0029 & -0.0287 & 0.0939 & 7.01(10) & $7.06(1)_{\rm e}(4)_{\rm t}$\tablenotemark[4] & -0.7\%\\
 $6p_{3/2}$--$5d_{3/2}$ & 4.0625 & -0.1465 & -0.7838 & -0.0054 & -0.0014 & -0.0129 & 0.0449 & 3.157(45) & $3.18(1)_{\rm e}(2)_{\rm t}$\tablenotemark[4]& -0.8\%\\
 $7p_{1/2}$--$5d_{3/2}$ & 4.0395 & 0.1075 & -2.2035 & -0.0185 & -0.0036 & -0.0427 & 0.1467 & 2.03(15) & 2.033(5)~\cite{Toh2019a}&-0.1\% \\
 $7p_{3/2}$--$5d_{3/2}$ & 1.6880 & 0.0478 & -0.9741 & -0.0068 & -0.0017 & -0.0182 & 0.0638 & 0.799(64) 
 & 0.799(5)\tablenotemark[4]&-0.01\%^\dagger\\
 &&&&&&&&& 0.795(4)\tablenotemark[4] & 0.5\%^\dagger\\
 $6p_{3/2}$--$5d_{5/2}$ & 12.1864 & -0.4350 & -2.2430 & -0.0201 & -0.0034 & -0.0401 & 0.1319 & 9.58(13) & 9.650(18)~\cite{Pucher2020} & -0.7\% \\
 $7p_{3/2}$--$5d_{5/2}$ & 5.0246 & 0.1407 & -2.7820 & -0.0253 & -0.0045 & -0.0528 & 0.1941 & 2.49(20) 
 & 2.493(15)\tablenotemark[4]&-0.01\%^\dagger \\
 &&&&&&&&& 2.481(11)\tablenotemark[4]&0.4\%^\dagger \\
\end{tabular}
\tablenotemark[1]{Average of Refs.~\cite{Tanner1992,Young1994a,RafTan98,*Rafac1999a,Derevianko2002a,Amini2003,
Bouloufa2007,Sell2011,*Patterson2015,Zhang2013,Gregoire2015}};
\tablenotemark[2]{Th.+Expt.~\cite{Safronova1999,Bennett1999}};
\tablenotemark[3]{Th.+Expt.~\cite{TohBeta2019,Bennett1999}};
\tablenotemark[4]{Th.+Expt.~(this work, see text)}.
\end{ruledtabular}
\end{table*}

The most important correlations that are included in coupled-cluster--type calculations, but not in the current version of our method, are ladder diagrams~\cite{DzubaLadder2008}.
Such diagrams are suppressed by an extra energy denominator corresponding to excitation from the core.
They can be important in some cases, e.g.,\ $d$ states,
and become appreciable for moderately-charged ions~\cite{RobertsActinides2013}.
In Ref.~\cite{DzubaLadder2008}, it was demonstrated that the ladder diagrams lead to a dramatic improvement in $d$-state energies, while not negatively impacting $s$ or $p$ energies.
The ladder diagram calculations are computationally intensive, and we do not include them in this work.
They are expected to contribute negligibly to E1 amplitudes, and most of this can be accounted for via the scaling.
A full inclusion of the ladder diagrams into the {\sc ampsci} correlation potential method and the subsequent affect on matrix elements will be considered in a coming work~\cite{Ladder2022}.

\subsection{Uncertainty estimate}\label{sec:uncert}

The dominant source of uncertainty in our calculations comes from the omission of certain correlation corrections, e.g., ladder diagrams.
We can estimate the impact of this by comparing the results calculated at different approximations.
To estimate the uncertainty due to missed Brueckner-type correlation effects, we first take an error term equal to the semi-empirical scaling correction ($\delta\lambda$ in the tables).
This correction stems from re-scaling the correlation potential, and as such, is mainly sensitive to correlation errors that are proportional to the correlation potential, $\Sigma$.
To account for errors which are not linear in $\Sigma$, we also add half the difference between scaled second-order and scaled all-orders ($\lambda\Sigma^{(\infty)}-\lambda\Sigma^{(2)}$) results (see Table~\ref{tab:E1-lambda}).
For the considered ions, where higher-order correlations are expected to be more important compared to neutral atoms, we instead take the full ($\lambda\Sigma^{(\infty)}-\lambda\Sigma^{(2)}$) difference on top of the $\delta\lambda$ contribution.

There are also the non-Brueckner correlation effects, SR and NS.
Based on the spread of these values between various levels of approximation, we estimate the uncertainty stemming from this contribution to be approximately 30\%.
We further add a conservative uncertainty of 50\% of the Breit contribution (Breit contributes negligibly to most E1 amplitudes, though is more important for the considered ions, and for transitions involving $d$-states).
Based on the results of our recent work~\cite{FairhallQED2022}, we take the uncertainty in the QED correction to be 25\%.
The final theory uncertainty is then taken as the sum in quadrature of all the above.

It is possible to calculate ratios between fine-structure pairs of matrix elements with very high precision. 
Such ratios differ from exact angular constants only due to relativistic effects, and depend weakly on correlations.
Theoretical ratios may be combined with experimental data to extract values for E1 matrix elements with high precision.
The uncertainty in these ratios is dominated by SR and we conservatively add 100\% of the combined SR+NS contribution on top of the above estimate.

The appropriateness of our method for determining the theory uncertainties is demonstrated by comparing with existing experimental results.
As shown below, the comparison is better than expected, indicating our uncertainties are conservative.
This gives us confidence in our calculations and uncertainties for the transitions where high-precision experiment is not yet available.

\section{Results and Comparison}\label{sec:res}

We performed high-precision calculations for E1 transitions between the low-lying $s$, $p$, and $d$ states of K, Ca\+ Rb, Sr\+, Cs, Ba\+, Fr, and Ra\+.
We find overall excellent agreement between our calculations and experiment, with typical deviations at the $0.1\%$ level.
The most extensive experimental data available is for Cs.
In Table~\ref{tab:E1-Cs} we present the full breakdown of contributions to the theoretical calculations for the Cs transitions, and comparison with experiment.
Complete tables of the results for the other systems are presented in Appendix~\ref{sec:A-E1}.
We present calculated ratios for several Cs transitions in Table~\ref{tab:Cs-ratios}, along with experimental values for comparison. The agreement with experiment is excellent.

\begin{table}
\caption{Ratios of reduced E1 matrix elements between the lowest few 
fine-structure pairs
of Cs as calculated in this work, and comparison with experiment.\label{tab:Cs-ratios}}
\begin{ruledtabular}
\begin{tabular}{llll}
 &  & \multicolumn{2}{c}{E1 Ratio: $|$A/B$|$} \\
\cline {3-4}
A & B &  Theory & \multicolumn{1}{c}{Expt.} \\
\hline
$6s$--$6p_{3/2}$ & $6s$--$6p_{1/2}$  &   1.40733(22)	&   1.4074(3)~\cite{RafTan98}\\ 
$6s$--$7p_{3/2}$ & $6s$--$7p_{1/2}$  &   2.070(35)	&   2.0646(26)~\cite{Damitz2019}\\ 
$7s$--$6p_{3/2}$ & $7s$--$6p_{1/2}$  &  1.5273(15)	&   1.5272(17)~\cite{Toh2019}\\ 
$7s$--$7p_{3/2}$ & $7s$--$7p_{1/2}$  &   1.38913(17)	&   \\ 
$5d_{3/2}$--$6p_{1/2}$ & $5d_{3/2}$--$6p_{3/2}$  &  2.2184(18)    &  \\ 
$5d_{3/2}$--$7p_{1/2}$ & $5d_{3/2}$--$7p_{3/2}$  &   2.535(20)	&   \\ 
$6p_{3/2}$--$5d_{5/2}$ & $6p_{3/2}$--$5d_{3/2}$  &   3.0330(16)	&   \\ 
$7p_{3/2}$--$5d_{5/2}$ & $7p_{3/2}$--$5d_{3/2}$  &   3.1222(92)	&   \\ 
\end{tabular}
\end{ruledtabular}
\end{table}

\begin{figure*}
\centering\tiny
\includegraphics[width=\linewidth]{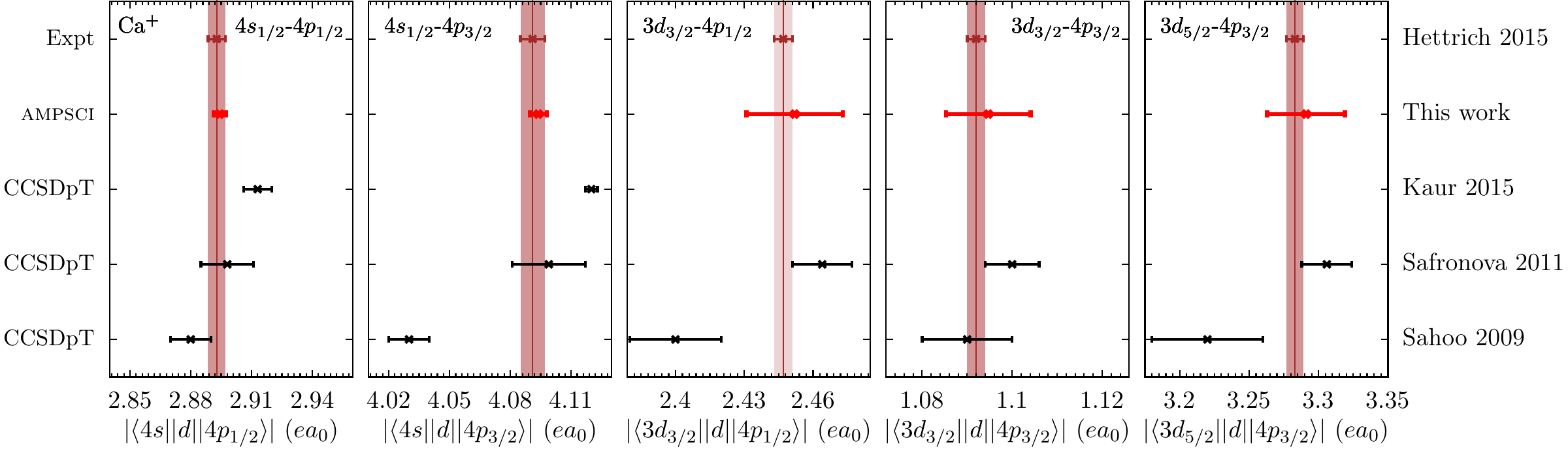}
\caption{\small Reduced E1 matrix elements (absolute values) for Ca\+ transitions.
The shaded regions are experimental values with $1\,\sigma$ uncertainties from Refs.~\cite{UDelAtomPortal,Hettrich2015,Ramm2013} (the lighter shaded region is the value extracted from experiment~\cite{Hettrich2015} in this work, see text for details). 
The red points are the \textsc{ampsci} calculations from this work, and the black points are other theory values: 
`Sahoo 2009'~\cite{Sahoo2009a},
`Safronova 2011'~\cite{Safronova2011}, 
and `Kaur 2015'~\cite{Kaur2015}.
\label{fig:Ca}}
\end{figure*}

\begin{figure*}
\centering\tiny
\includegraphics[width=0.495\linewidth]{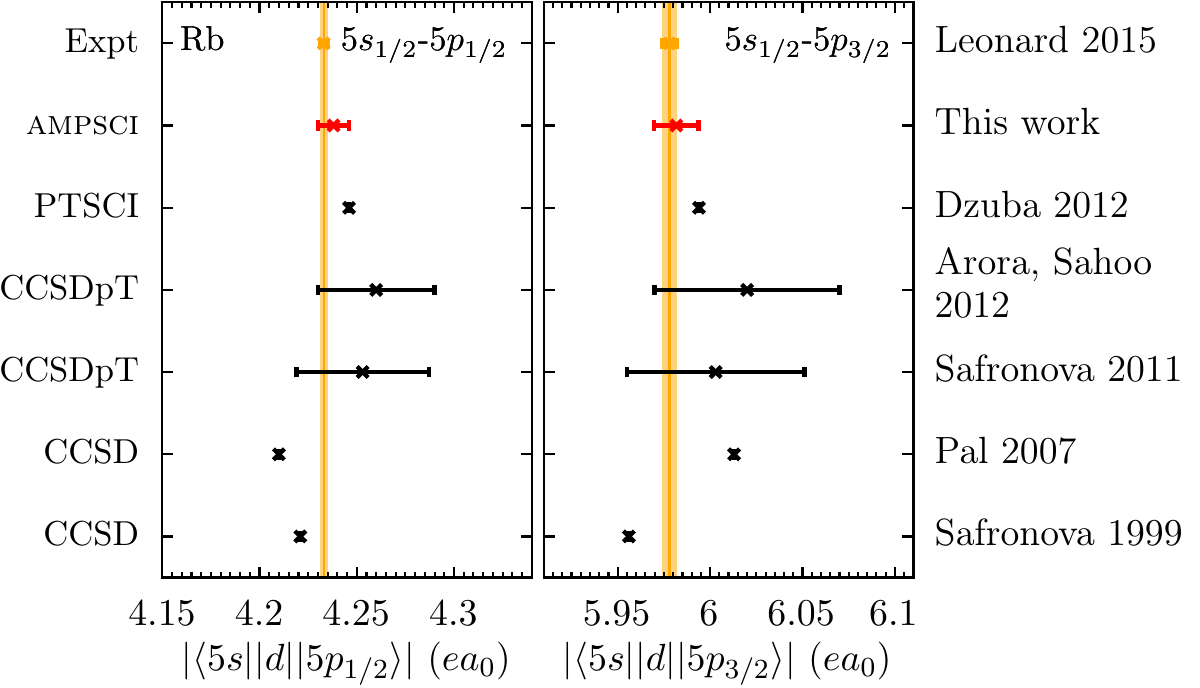}
\includegraphics[width=0.495\linewidth]{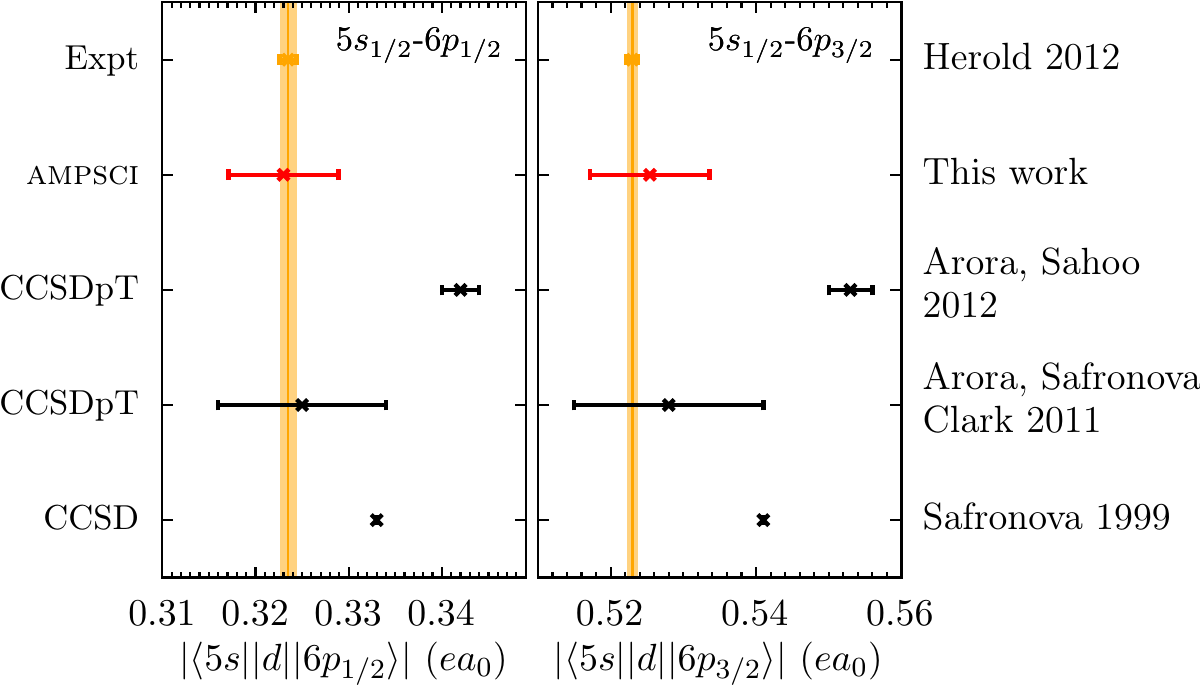}
\caption{\small 
Rb transitions.
Experimental values are from 
`Leonard 2015'~\cite{Leonard2015,UDelAtomPortal}
and `Herold 2012'~\cite{Herold2012};
other theory values are from 
`Safronova 1999'~\cite{Safronova1999}, 
`Pal 2007'~\cite{Pal2007}, 
`Safronova 2011'~\cite{safronova11a}, 
`Arora, Safronova, Clark 2011'~\cite{AroSafCla11,Herold2012}, 
`Arora, Sahoo 2012'~\cite{Arora2012}, and 
`Dzuba 2012'~\cite{OurRb2012}.
\label{fig:Rb5s}}
\end{figure*}

We discuss the comparisons with experiment and other theory for a few instructive cases in the following sections.
To aid in the comparison between results, we present plots of our calculations alongside known experimental values and other theory results. 
We only plot those transitions for which there is sufficient other theory or experimental results to compare.
In the figures, we refer to the method used for the theoretical calculations:\
CCSD is the coupled-cluster approach including single and double excitations (see, e.g., Ref.~\cite{Safronova2008});
CCSDpT is as above with certain valence triple excitations included perturbatively~\cite{Blundell1989,*Dame1991,SafDerJoh98,Safronova1999};
CCSDvT is with valence triple excitations fully included in the CC formulation~\cite{Porsev2006};
CCSDT is with both valence and core triples included~\cite{Porsev2021b};
PTSCI is the perturbation theory in the screened Coulomb interaction (correlation potential method)~\cite{DzubaCPM1988pla,DzubaCPM1989plaEn};
and {\sc ampsci} refers to our implementation of the PTSCI method as described in the theory section.

\begin{figure*}
\centering\tiny
\includegraphics[width=\linewidth]{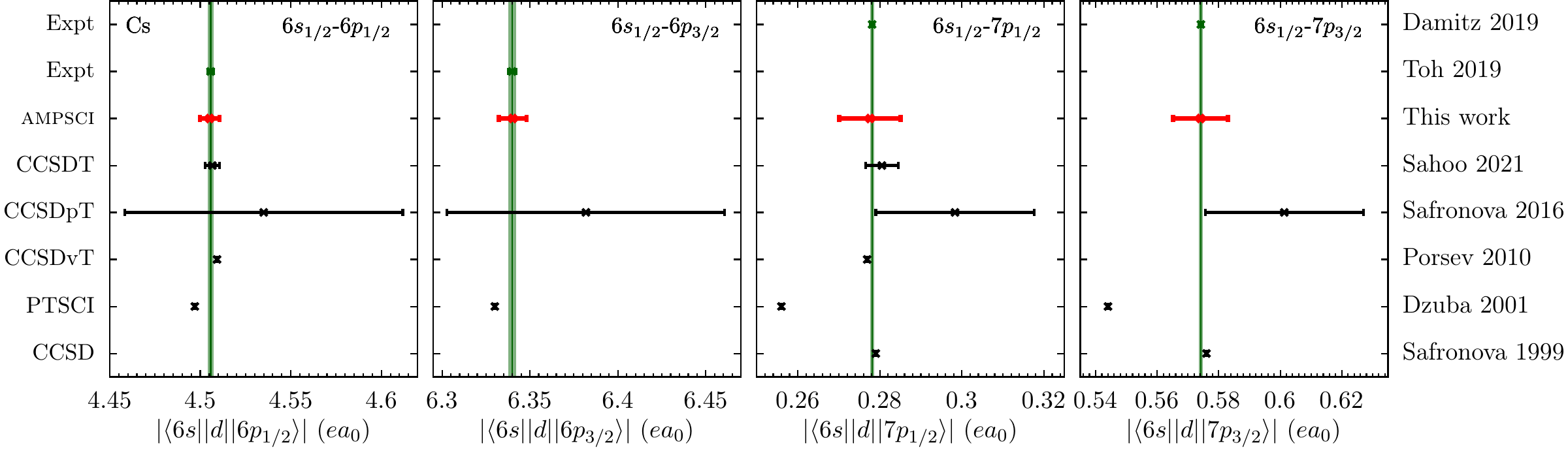}
\caption{\small Cs $6s$ transitions.
Experimental values are from 
`Toh 2019'~\cite{TohBeta2019} and
`Damitz 2019'~\cite{Damitz2019};
theory values are from 
`Safronova 1999'~\cite{Safronova1999},
`Dzuba 2001'~\cite{DzubaPNCsd2001,GingesCs2002},
`Porsev 2010'~\cite{Porsev2010},
`Safronova 2016'~\cite{Safronova2016b}, and
`Sahoo 2021'~\cite{Sahoo2021}.
\label{fig:Cs6s}}
\end{figure*}

\subsubsection{Calcium ion}

For Ca\+, our calculations are in excellent agreement with experiment, and with the calculations of 
Ref.~\cite{Safronova2011}, 
as shown in Fig.~\ref{fig:Ca}.
However, there is strong tension between our results and those of Refs.~\cite{Sahoo2009a,Kaur2015}.
The $4s$--$4p_{3/2}$ result of Ref.~\cite{Sahoo2009a} disagrees with experiment at the $5\,\sigma$ level, and the $4s$--$4p_{j}$ and $4s$--$4p_{3/2}$ results of Ref.~\cite{Kaur2015} disagree at the $3\,\sigma$ and $5\,\sigma$ levels, respectively.

As discussed above, we may combine theoretical ratios with experimental data to extract experimental values for dipole amplitudes.
We calculate the ratio
\begin{equation}
\frac{|\bra{4p_{1/2}}| d|\ket{3d_{3/2}}|}{|\bra{4p_{3/2}}| d|\ket{3d_{3/2}}|}
=2.2404(3).
\end{equation}
Combining this with the experimental 
$4p_{3/2}$--$3d_{3/2}$ value of 1.092(2)\,$ea_0$~\cite{UDelAtomPortal,Hettrich2015,Ramm2013}, we extract:
\begin{equation}
|\bra{4p_{1/2}}| d|\ket{3d_{3/2}}| = 2.447(4)\,ea_0,
\end{equation}
where the uncertainty is dominated by experiment.

\subsubsection{Rubidium}

For the Rb transitions we find excellent agreement with experiment, and good agreement with most other high-precision calculations, see Fig.~\ref{fig:Rb5s}.
For the $5s$--$6p_j$ transitions, our calculations agree extremely well with the recent experimental results of Harold {\sl et al.}~\cite{Herold2012}, and with the CCSDpT calculations also from Ref.~\cite{Herold2012}.
However, we disagree significantly with the CCSDpT results of 
Ref.~\cite{Arora2012}, which also disagree with experiment at the $9\,\sigma$ level.
The ratio of the lowest $s$-$p$ matrix elements was also measured recently with high precision~\cite{Leonard2015,LeonardErratum2017}, and is in perfect agreement with our calculation:
\begin{equation}\label{eq:RbRatio}
\frac{|\bra{5s}| d|\ket{5p_{3/2}}|}{|\bra{5s}| d|\ket{5p_{1/2}}|}
=
\begin{cases}
1.41141(9)& {\rm Theory}\\
1.41144(1) & \text{Expt.~\cite{Leonard2015,LeonardErratum2017}}
.
\end{cases}
\end{equation}

\subsubsection{Strontium ion}

For Sr\+, our calculations are in good agreement with experiment and with previous theory calculations~\cite{Jiang2009,Safronova2011a,Kaur2015,Jiang2016,Kaur2021}.
The experimental amplitudes were determined in this work by combining high-precision measurements of branching ratios from Ref.~\cite{Zhang2016} with lifetime measurements from Ref.~\cite{Pinnington1995}. 
The experimental error ($\sim$\,1\%) is dominated by uncertainty in the lifetimes, and is larger than for the other atoms considered here.
We may, however, directly compare our calculations to the high-precision experimental branching fractions from Ref.~\cite{Zhang2016}, as shown in Table~\ref{tab:sr-branch}.
The comparison shows excellent agreement with theory, again indicating that our theoretical uncertainties are conservative.

\begin{table}
\caption{Branching fractions for decays from the Sr\+ $5p$ states, and comparison with experiment.
The notation used in Ref.~\cite{Zhang2016} is repeated here for convenience.
\label{tab:sr-branch}}
\begin{ruledtabular}
\begin{tabular}{lllll}
Upper & Lower & Label~\cite{Zhang2016} &Theory & \multicolumn{1}{c}{Expt.~\cite{Zhang2016}} \\
\hline
$5p_{1/2}$ & $5s$       & $p$   & 0.94496(60) & 0.94498(8) \\
         & $5d_{3/2}$ & $1-p$ & 0.05504(60) & 0.05502(8) \\
$5p_{3/2}$ & $5s$       & $q$   & 0.94065(8) & 0.9406(2)  \\
         & $5d_{3/2}$ & $r$   & 0.00630(15)  & 0.0063(3)  \\
         & $5d_{5/2}$ & $s=1-q-r$   & 0.05305(12) & 0.0531(2) \\
\end{tabular}
\end{ruledtabular}
\end{table}





\begin{figure*}
\centering\tiny
\includegraphics[width=\linewidth]{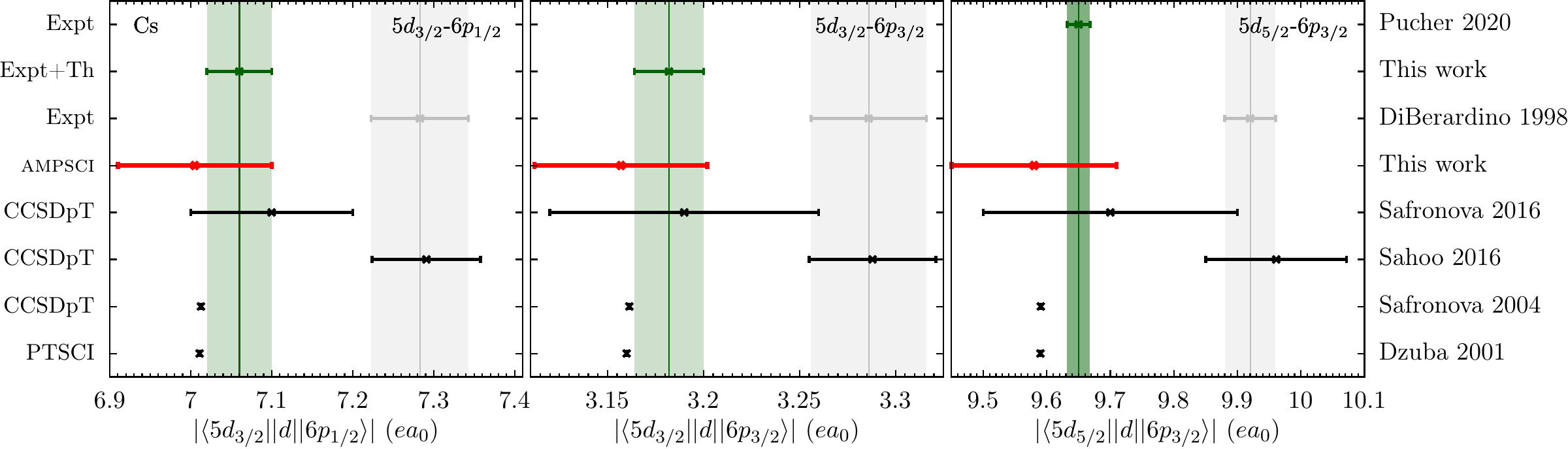}
\caption{\small Cs $5d$ transitions.
Experimental values are from 
`DiBerardino 1998'~\cite{DiBerardino1998}, and 
`Pucher 2020'~\cite{Pucher2020};
other theory values are from 
`Dzuba 2001'~\cite{DzubaPNCsd2001,Robertssd2014},
`Safronova 2004'~\cite{Safronova2004b},
`Sahoo 2016'~\cite{Sahoo2016}, and
`Safronova 2016'~\cite{Safronova2016b}.
The `Expt+Th' values were determined in this work (see text).
The experimental results from Ref.~\cite{DiBerardino1998} (shown in grey) are inconsistent with measurements of polarizabilities (see text and Refs.~\cite{Pucher2020,Safronova2004b}).
The calculations of Ref.~\cite{Sahoo2016} agrees almost exactly with the previous experiment~\cite{DiBerardino1998},
though disagrees (by 2.8$\,\sigma$) with more accurate subsequent measurements~\cite{Pucher2020}.
\label{fig:Cs5d}}
\end{figure*}

\subsubsection{Cesium}

The agreement for transitions involving the lowest $s$ and $p$ states in Cs is particularly good, as shown  in Fig.~\ref{fig:Cs6s}.
There is also very good agreement between theory calculations of different groups.
One of the differences between theory values is that between our results and those of Ref.~\cite{DzubaPNCsd2001} for the $6s$--$7p_j$ transitions.
As can be seen from Table~\ref{tab:E1-Cs}, the large structure radiation correction for these transitions has the same sign and magnitude to account for the difference.
We also note that the apparent large difference is also due to the smallness of the matrix element; the absolute difference is small, $\simeq$\,$0.03\,ea_0$.

A new measurement of the Cs $5d_{5/2}$ lifetime has been reported by Pucher {\sl et al.}~\cite{Pucher2020}, finding $\tau=1353(5)\,{\rm ns}$.
This measurement is in disagreement with a much earlier measurement of DiBerardino {\sl et al.}~\cite{DiBerardino1998}, $\tau=1281(9)\,{\rm ns}$.
However, it resolves tensions between experimental determinations of Cs lifetimes and polarizabilities, as predicted by Safronova {\sl et al.}~\cite{Safronova2004b} (see detailed discussions in Refs.~\cite{Pucher2020,Safronova2004b}).
There is a single E1 decay pathway for the $5d_{5/2}$ state, allowing an unambiguous determination for the $5d_{5/2}$--$5p_{3/2}$ matrix element:
\begin{equation}
|\bra{5d_{5/2}}|d|\ket{5p_{3/2}}|=9.650(18)\,ea_0.
\end{equation}
This is in excellent agreement with our calculation, and that of Safronova {\sl et al.}~\cite{Safronova2004b,Safronova2016b}, but is in strong disagreement with the calculation of Sahoo~\cite{Sahoo2016} (at the $3\,\sigma$ level).
This is shown in Fig.~\ref{fig:Cs5d}.

\begin{figure*}
\centering\tiny
\includegraphics[width=0.495\linewidth]{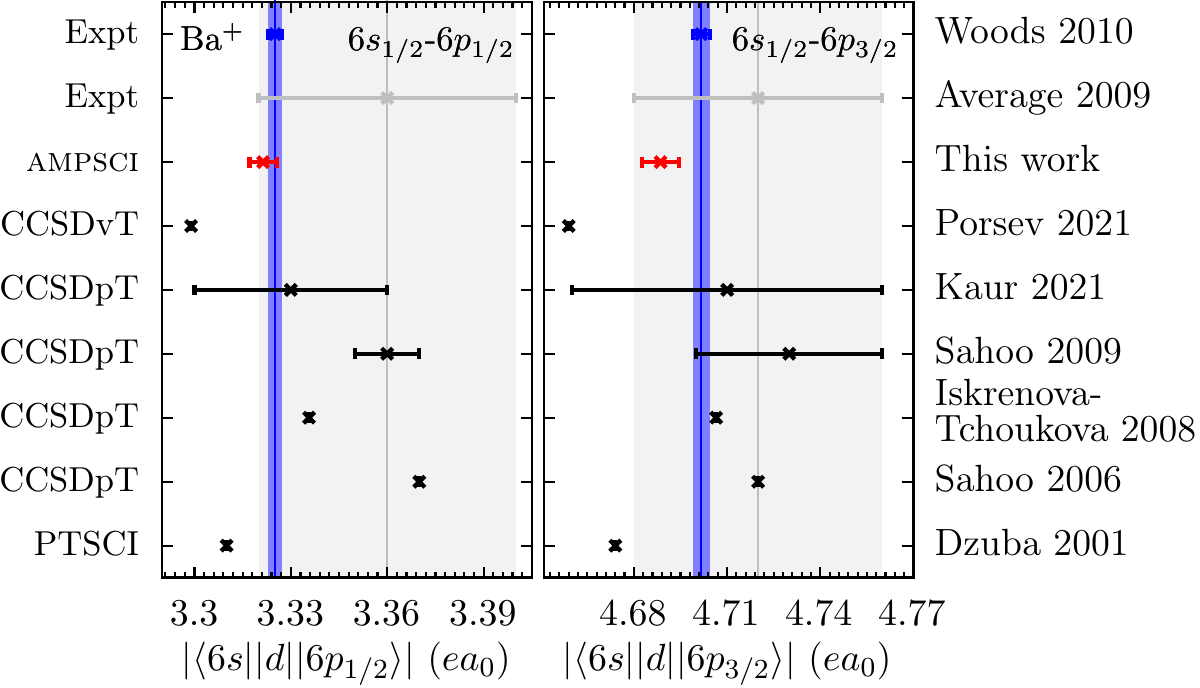}
\includegraphics[width=0.495\linewidth]{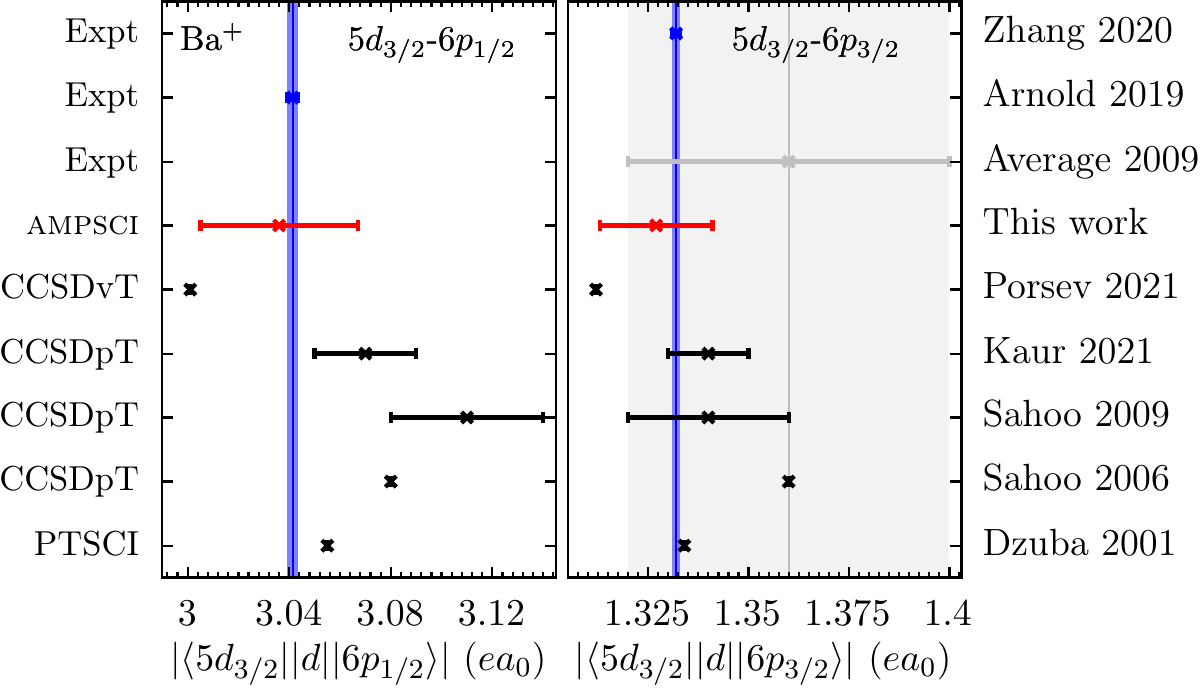}
\caption{\small Ba\+ transitions.
Experimental values are from Ref.~\cite{UDelAtomPortal} (`Woods 2010'~\cite{Woods2010},
`Arnold 2019'~\cite{Arnold2019},
`Zhang 2020'~\cite{Zhang2020a}); 
theory values are from
`Dzuba 2001'~\cite{DzubaPNCsd2001},
`Sahoo 2006'~\cite{Sahoo2006a},
`Iskrenova-Tchoukova 2008'~\cite{Iskrenova-Tchoukova2008},
`Sahoo 2009'~\cite{Sahoo2009},
`Kaur 2021'~\cite{Kaur2021},
`Porsev 2021'~\cite{Porsev2021};
`Average 2009' is average of experiments~\cite{Kastberg1993,Gallagher1967,Klose2002,Davidson1992a,Kurz2008} as quoted in Ref.~\cite{Sahoo2009}.
\label{fig:ba}}
\end{figure*}

Lifetimes for the $7p_{1/2}$ and $7p_{3/2}$ states were determined recently by Toh {\sl et al.}~\cite{Toh2019a}, finding 165.21(29)\,ns and 137.54(16)\,ns, respectively.
Combining the $7p_{1/2}$ lifetime with known E1 matrix elements, the authors determined the E1 matrix element for the $7p_{1/2}$--$5d_{3/2}$ transition with high precision. 
The resulting value, 2.033(5)\,$ea_0$, is in excellent agreement with our calculations.

Since the $7p_{3/2}$ state has four E1 decay pathways (to $6s$, $7s$, $5d_{3/2}$, and $5d_{5/2}$), in order to determine the $7p_{3/2}$--$5d_{j}$ matrix elements, one requires accurate determinations for the $6s$--$7p_{3/2}$ and $7s$--$7p_{3/2}$ matrix elements, and the ratio between the $7p_{3/2}$--$5d_{j}$ matrix elements.
The $6s$--$7p_{3/2}$ matrix element is known to high precision~\cite{Damitz2019}, though there is some ambiguity for the $7s$--$7p_{3/2}$ case.

There are two values for each of the $7s$--$7p_{j}$ matrix elements, each determined from transition polarizabilities measured in Ref.~\cite{Bennett1999} in combination with the theoretical ratio of $7s$--$7p_{j}$ matrix elements.
The values for $7s$--$7p_{3/2}$ are 
14.320(20)\,$ea_0$~\cite{UDelAtomPortal,Safronova1999} and
14.344(7)\,$ea_0$~\cite{TohBeta2019}, which disagree at the $1\,\sigma$ level, despite being derived from the same experiment~\cite{Bennett1999} (and similar for $7s$-$7p_{1/2}$).
The former agrees more closely with theoretical predictions~\cite{Safronova1999,DzubaPNCsd2001,GingesCs2002,Porsev2010,Safronova2016b,Sahoo2021}, though the latter has smaller uncertainty.
We also note that the lifetime measurement for $7p_{3/2}$ state~\cite{Toh2019a} appears to favor a smaller magnitude for the $7s$-$7p_j$ matrix elements.
Due to this ambiguity, we don't include these matrix elements in our uncertainty analysis. 
The $7s$--$7p_{1/2}$ matrix element is very important for calculations of atomic parity violation (see, e.g., Ref.~\cite{GingesCs2002}), so this warrants further study.

Combining the above matrix elements with the ratio from Table~\ref{tab:Cs-ratios} allows us to extract high-precision values for the $7p_{3/2}$--$5d_j$ matrix elements. 
Using the more precise $7s$--$7p_{3/2}$ value [14.344(7)\,$ea_0$~\cite{TohBeta2019}], we extract
\begin{align}
|\bra{5d_{3/2}}|d|\ket{7p_{3/2}}|&=0.795(4)\,ea_0,\\
|\bra{5d_{5/2}}|d|\ket{7p_{3/2}}|&=2.481(11)\,ea_0.
\end{align}
(The other value leads to
0.799(5) and 2.493(15), respectively.)
The uncertainties are dominated by experiment.

We may also use our calculated ratios along with the above-determined
$5p_{3/2}$--$5d_{5/2}$ matrix element to determine the two $6p_j$--$5d_{3/2}$ matrix elements:
\begin{align}
|\bra{5d_{3/2}}|d|\ket{6p_{1/2}}|&=7.06(1)_{\rm ex}(4)_{\rm th}\,ea_0,\\
|\bra{5d_{3/2}}|d|\ket{6p_{3/2}}|&=3.182(6)_{\rm ex}(17)_{\rm th}\,ea_0.
\end{align}
Unlike in the previous case, this uncertainty is dominated by theory (from the ratios in Table~\ref{tab:Cs-ratios}),
though it is smaller than for the direct calculations.
These are in disagreement with those derived from the lifetime measurements of Ref.~\cite{DiBerardino1998}, however, are consistent with polarizabilities~\cite{Safronova2004b,Safronova2016b}, and agree well with our calculations and those of 
Safronova {\sl et al.}
Ref.~\cite{Safronova2016b} (see Fig.~\ref{fig:Cs5d}).

It is worth remarking that in several cases newer more precise experiments have indicated better agreement with theory than was previously determined. 
For example, consider the following quote from 
Dzuba {\sl et al.}~\cite{DzubaPNCsd2001}:
{\it ``The 5$d$-6$p$ [E1 matrix elements] for Cs have poor accuracy, deviating from experiment by about 4\%. This is indicative of the poor calculation of $d$ states''}; in fact, the $5d_{3/2}$--$6p_{3/2}$ calculation from Ref.~\cite{DzubaPNCsd2001} agrees very well with the new experimental value~\cite{Pucher2020}, deviating by just $0.6\%$.
This has important implications for the assumed accuracy of subsequent atomic calculations.

\begin{figure*}
\centering\tiny
\includegraphics[width=\linewidth]{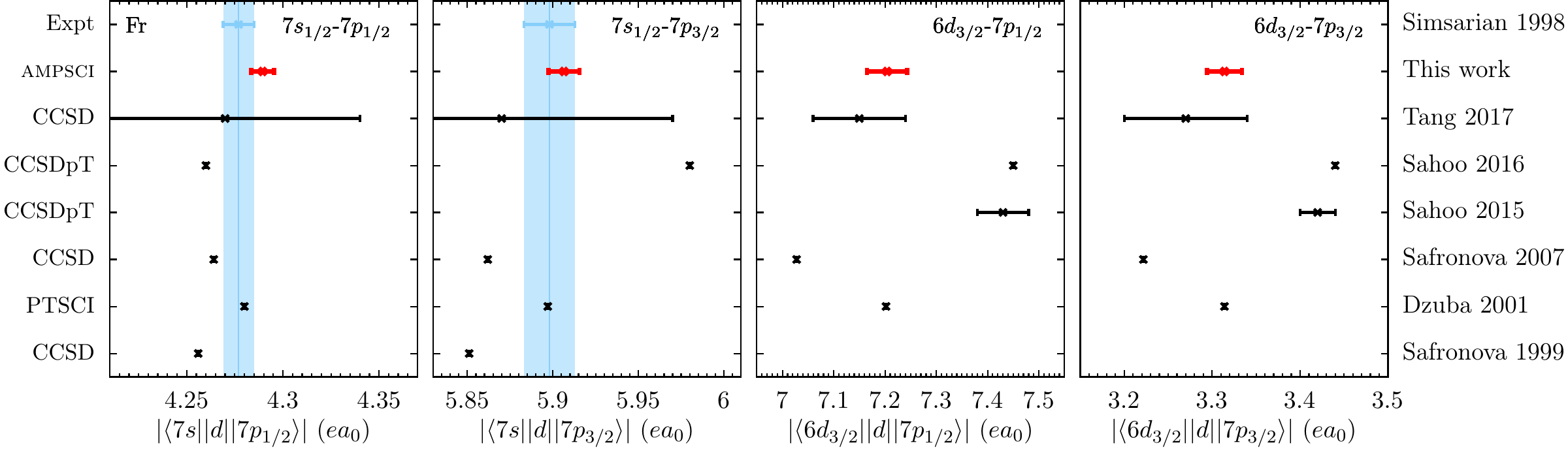}
\caption{\small Francium transitions.
Experimental values are from Refs.~\cite{UDelAtomPortal,Simsarian1998}; other theory values are from 
`Safronova 1999'~\cite{Safronova1999},
`Dzuba 2001'~\cite{DzubaPNCsd2001,RobertsActinides2013},
`Safronova 2007'~\cite{Safronova2007},
`Sahoo 2015'~\cite{Sahoo2015},
`Sahoo 2016'~\cite{Sahoo2016a},
`Tang 2017'~\cite{Tang2017}.
For the $p$-$d$ transitions there is a large spread in the theory values.
Based on the $p$-$d$ results in Cs, Ba\+, and Ra\+, we expect our calculations to be the most accurate.
\label{fig:fr}}
\end{figure*}

\subsubsection{Barium ion}

There are very precise recent experimental data available for several Ba\+ transitions.
With the exception of the $6s$--$6p_{3/2}$ transition (discussed below), our calculations agree exceptionally well with experiment, as shown in Fig.~\ref{fig:ba}.
There is also fair agreement among the results of most other calculations, though the spread in values is larger than for Cs.
Detailed calculations were performed for Ba\+ recently in Ref.~\cite{Porsev2021}, where particular attention was paid to the role of triple excitations.
Those calculations are in good agreement with ours, though our calculations lie closer to the experimental values.
Our calculations are in significant tension with those of 
Ref.~\cite{Sahoo2009}. 
Note that the $6s$--$6p_{1/2}$ calculation of Ref.~\cite{Sahoo2009} agreed precisely with the midpoint of the best experimental value that was available at the time of the calculation~\cite{Klose2002,Davidson1992a}, but disagrees (by $3.4\,\sigma$ sigma) with the more accurate subsequent measurement of Ref.~\cite{Woods2010}.

For the $6s$--$6p_{3/2}$, our calculation disagrees with experiment at the 2\,$\sigma$ level, the largest disagreement for all our calculations. 
Even here, the absolute agreement is very good -- the midpoints of the theory and experiment differ by only $0.01\,ea_0$ (or $0.3\%)$ -- however, both the experimental and estimated theory uncertainties are small.

We also calculate the ratio of the $6s$-$6p_j$ matrix elements, and find a $2\,\sigma$ tension with experiment:
\begin{equation}\label{eq:R-ba}
\frac{|\bra{6s}| d|\ket{6p_{3/2}}|}{|\bra{6s}| d|\ket{6p_{1/2}}|}
=
\begin{cases}
1.4116(2) & {\rm Theory}\\
1.4140(12) & \text{Expt.~\cite{Woods2010}}.
\end{cases}
\end{equation}
Since the theory accuracy is expected to be high this is an important discrepancy, as noted previously~\cite{Arnold2019}.
Ratios derived from other high-precision theory --
1.412~\cite{DzubaPNCsd2001},
1.4109(2)~\cite{Iskrenova-Tchoukova2008,Arnold2019}, 
1.411~\cite{Kaur2021}, 
and 1.412~\cite{Porsev2021} -- also disagree with the measured value, though are in agreement with our calculation.
This may indicate that the experimental uncertainties for the $6s$--$6p_j$ transitions have been underestimated.
(The ratio 1.40 from Ref.~\cite{Sahoo2006a} disagrees both with experiment and other theory.) 

\subsubsection{Francium}

\begin{table}
\caption{Ratios of reduced E1 matrix elements between the lowest few states of Fr.\label{tab:Fr-ratios}}
\begin{ruledtabular}
\begin{tabular}{llll}
 &  & \multicolumn{2}{c}{E1 Ratio: $|$A/B$|$} \\
\cline {3-4}
A & B & Theory & \multicolumn{1}{c}{Expt.} \\
\hline
$7s$--$7p_{3/2}$ & $7s$--$7p_{1/2}$  &   1.3770(7)	&  1.379(4)~\cite{Simsarian1998} \\ %
$8s$--$7p_{3/2}$ & $8s$--$7p_{1/2}$  &   1.762(7)	&   \\ %
$6d_{3/2}$--$7p_{1/2}$ & $6d_{3/2}$--$7p_{3/2}$  &   2.174(2)  &  \\ 
$7p_{3/2}$--$6d_{5/2}$ & $7p_{3/2}$--$6d_{3/2}$  &  3.072(3)	&  \\ 
\end{tabular}
\end{ruledtabular}
\end{table}

\begin{figure*}
\centering\tiny
\includegraphics[width=\linewidth]{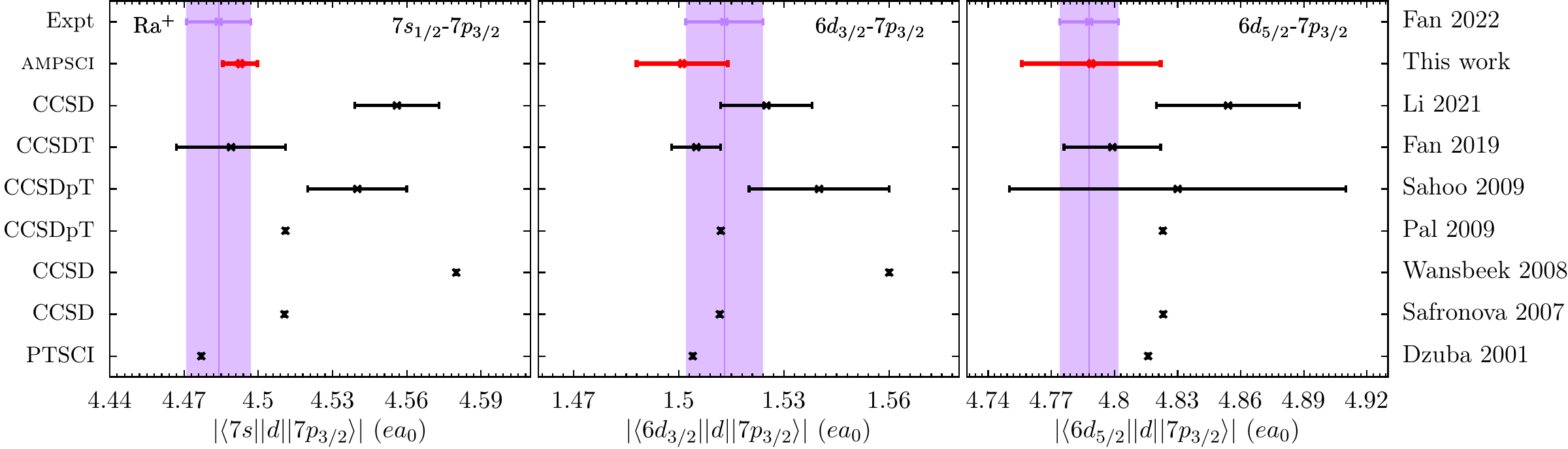}
\caption{\small Ra\+ $7p_{3/2}$ transitions.
Experimental values are from Ref.~\cite{Fan2022}; other theory values are from 
`Dzuba 2001'~\cite{DzubaPNCsd2001,RobertsActinides2013},
`Safronova 2007'~\cite{Safronova2007},
`Wansbeek 2008'~\cite{Wansbeek2008},
`Pal 2009'~\cite{Pal2009},
`Sahoo 2009'~\cite{Sahoo2009},
`Fan 2019'~\cite{Fan2019a}, and `Li 2021'~\cite{LiRa2021}.
Overall, there is good agreement between results, except for those from Refs.~\cite{Wansbeek2008,Sahoo2009,LiRa2021}.
\label{fig:ra+p32}}
\end{figure*}

For Fr, there is good agreement between most theory results and experiment, as shown in Fig.~\ref{fig:fr}.
However, there are a few significant outliers, and limited experimental data. 
There is tension between our calculations and those of Refs.~\cite{Sahoo2015,Sahoo2016a} for the $7s$--$7p_{3/2}$ and $7p$--$6d$ transitions, significantly larger disagreement than expected from the claimed uncertainties.
Although there is insufficient high-precision experimental data for Fr available to distinguish between the theory results for the $p$-$d$ transitions, based on the $7s$--$7p_{3/2}$ transition, as well as the $p$-$d$ results in Cs [Fig.~\ref{fig:Cs5d}], Ba\+ [Fig.~\ref{fig:ba}], and Ra\+ [Fig.~\ref{fig:ra+p32}], we expect our calculations are most accurate.

Calculations of E1 ratios for Fr are presented in Table~\ref{tab:Fr-ratios}.
Combining these with the measured 8s lifetime, 53.3(4)\,ns~\cite{Gomez2005}, we extract:
\begin{align}
|\bra{8s}|d|\ket{7p_{1/2}}|&=4.234(17)_{\rm ex}(10)_{\rm th}\,ea_0,\\
|\bra{8s}|d|\ket{7p_{3/2}}|&=7.460(31)_{\rm ex}(12)_{\rm th}\,ea_0,
\end{align}
where the uncertainties are dominated by experiment.
These are in excellent agreement with our calculations.

\subsubsection{Radium ion}

The lifetime of the 
$7p_{3/2}$
state in Ra\+ was measured very recently~\cite{Fan2022},
and found to be
$
\tau = 4.78(3)\,{\rm ns}.
$
By combining this measurement with data for branching ratios and frequencies from Refs.~\cite{Fan2019a,Holliman2020},
the authors of Ref.~\cite{Fan2022} were able to extract accurate determinations of the $7s$--$7p_{3/2}$, $6d_{3/2}$--$7p_{3/2}$, and $6d_{5/2}$--$7p_{3/2}$ E1 matrix elements.
These determinations are in excellent agreement with our calculations (we note that our calculations were completed before Ref.~\cite{Fan2022} was posted on arXiv).

\begin{figure}
\centering
\includegraphics[width=\linewidth]{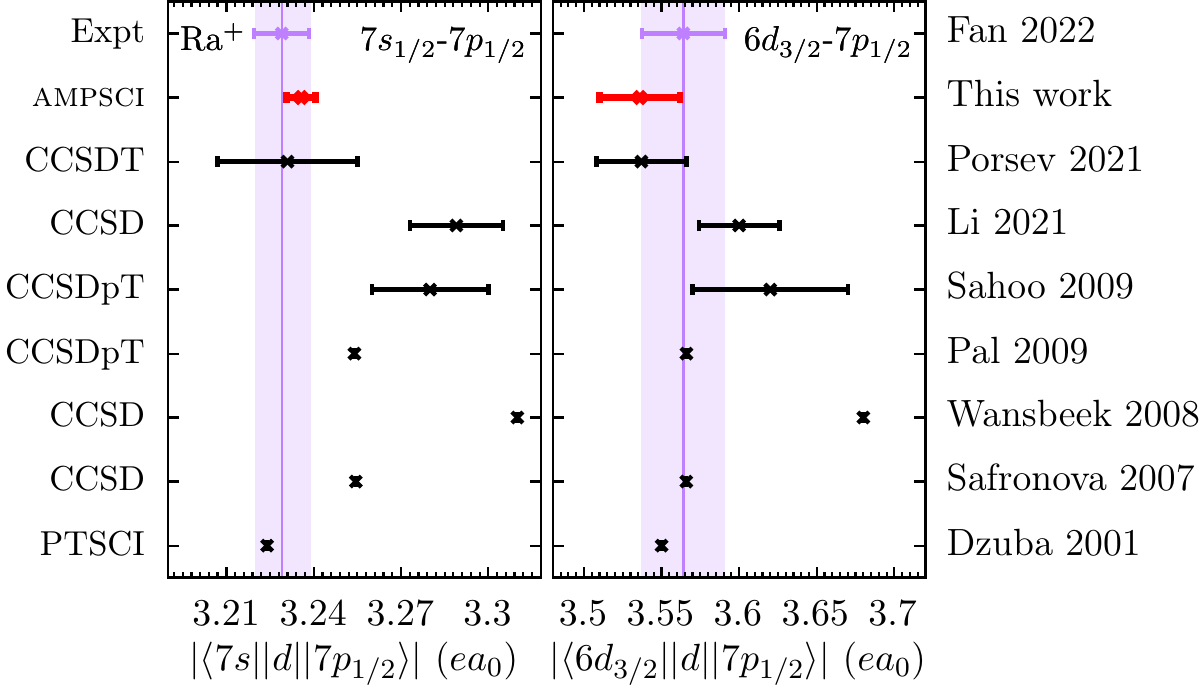}
\caption{\small Ra\+ $7p_{1/2}$ transitions.
The light shaded regions show the values extracted in this work from experiment~\cite{Fan2022}.
Other theory values are from 
`Dzuba 2001'~\cite{DzubaPNCsd2001},
`Safronova 2007'~\cite{Safronova2007},
`Wansbeek 2008'~\cite{Wansbeek2008},
`Pal 2009'~\cite{Pal2009},
`Sahoo 2009'~\cite{Sahoo2009}, 
`Li 2021'~\cite{LiRa2021},
and
`Porsev 2021'~\cite{[][{, (private communication via Ref.~\cite{UDelAtomPortal}).}]PorsevPrivate,UDelAtomPortal}.
\label{fig:ra+}}
\end{figure}

Our calculations are also in good agreement with most previous high-precision calculations, shown in Figs.~\ref{fig:ra+p32} and \ref{fig:ra+}.
The calculations of Ref.~\cite{Wansbeek2008} differ substantially from ours and from the experimental results. 
This may be due to the relatively simple CCSD method used there, though the CCSD values from Ref.~\cite{Safronova2007} agree well.
The CCSDpT results of 
Ref.~\cite{Sahoo2009} 
disagree significantly with the other calculations (including other CC calculations), and their result for the $7s$--$7p_{3/2}$ disagrees significantly with the experimental result (by 2.3$\,\sigma$).

Finally, we also calculate ratios of E1 reduced matrix elements between some of the lowest states of Ra\+, shown in Table~\ref{tab:Ra+-ratios}.
The $7p_{3/2}$--$6d_j$ ratio was determined experimentally in Ref.~\cite{Fan2019a} (using wavelengths measured in Ref.~\cite{Holliman2020}).
This only just agrees with the calculated ratio, with a $1\,\sigma$ deviation.
Our calculated ratio agrees nearly perfectly with another theory value of 3.189(10) (CCSDT), also presented in Ref.~\cite{Fan2019a}.
We can combine our calculated ratio with the $7p_{3/2}$ matrix elements 
from Ref.~\cite{Fan2022} to extract accurate values for the $7s_{1/2}$--${7p_{1/2}}$ and $6d_{3/2}$--$7p_{1/2}$ matrix elements.
 We find:
\begin{align}
|\bra{7s_{1/2}}|d|\ket{7p_{1/2}}|&=3.229(9)_{\rm ex}(1)_{\rm th}\,ea_0,\\
|\bra{6d_{3/2}}|d|\ket{7p_{1/2}}|&=3.564(26)_{\rm ex}(7)_{\rm th}\,ea_0,
\end{align}
where the uncertainties are dominated by experiment.

\section{Discussion}

Of the 46 E1 transitions we consider here that have unambiguous high-precision measurements, only two of our theoretical values lie outside the 1$\,\sigma$ deviation from experiment (combining theory and experimental uncertainties). 
This is much better than statistically expected, indicating that our theory uncertainties are conservative.

\begin{table}
\caption{Ratios of reduced E1 matrix elements between the lowest few states of Ra\+.\label{tab:Ra+-ratios}}
\begin{ruledtabular}
\begin{tabular}{llll}
 &  & \multicolumn{2}{c}{E1 Ratio: $|$A/B$|$} \\
\cline {3-4}
A & B & Theory & \multicolumn{1}{c}{Expt.} \\
\hline
$7s$--$7p_{3/2}$ & $7s$--$7p_{1/2}$  &   1.3885(2)	&   \\ %
$8s$--$7p_{3/2}$ & $8s$--$7p_{1/2}$  &   1.843(9)	&   \\ %
$6d_{3/2}$--$7p_{1/2}$ & $6d_{3/2}$--$7p_{3/2}$  &   2.356(4)  &  \\ 
$7p_{3/2}$--$6d_{5/2}$ & $7p_{3/2}$--$6d_{3/2}$  &  3.190(10)	&  3.164(21)~\cite{Fan2019a} \\ 
\end{tabular}
\end{ruledtabular}
\end{table}

The biggest discrepancy occurs for the $6s$--$6p_{3/2}$ transition in Ba\+, which deviates by 2$\,\sigma$ (Fig.~\ref{fig:ba}).
A single $2\,\sigma$ disagreement in 46 cases is about what is statistically expected.
Note that the absolute agreement here is still very good:\ the midpoints differ by only $0.01\,ea_0$, though the claimed uncertainties are even smaller.
We also note that a surprising disagreement between theory and experiment for the ratio of amplitudes in Ba\+ [Eq.~\eqref{eq:R-ba}] may 
indicate the experimental uncertainties have been underestimated, as discussed above.
The other discrepancy occurs for the $7s$--$7p_{1/2}$ transition in Fr, which disagrees by $1.3\,\sigma$ (Fig.~\ref{fig:fr}).
The absolute disagreement is small here also, at just $0.01\,ea_0$.

While within the uncertainties, the largest relative difference between the theory and experimental midpoints occurs for the $6d_{3/2}$--$7p_{3/2}$ transition in Ra\+ ($0.8\%$, Fig.~\ref{fig:ra+p32}).
This is due to the smallness of the matrix element, the absolute difference is only $0.01\,ea_0$. 

Overall, we find excellent agreement between our calculations and those based on the CC method, provided that triple excitations are included in those calculations.
There are a few notable exceptions as discussed above.
One observation is that there is sometimes a large spread in results calculated using CC methods (e.g., see Figs.~\ref{fig:Ca}, 
\ref{fig:Cs5d}, \ref{fig:ba}).
This may be due to the sensitivity of the method to basis choices and to the details of including triple excitations.
Further, recent high-precision measurements have demonstrated that, in a few cases, calculations based on the CC method have significantly lower accuracy than originally claimed
(e.g., see Figs.~\ref{fig:Ca},
\ref{fig:Rb5s},
\ref{fig:Cs5d},
\ref{fig:ba},
\ref{fig:ra+p32}).
This highlights the need for a robust and reliable method for determining theoretical uncertainties.

In other cases, however, the agreement with recent experiment is found to be substantially better than originally expected.
For example, our calculations and the CCSDT calculations of Ref.~\cite{PorsevPrivate,Fan2019a} for Ra\+ were performed before the measurements of Ref.~\cite{Fan2022}, and agree exceptionally well, see Fig.~\ref{fig:ra+p32}.
Also, the accuracy of the PTSCI calculations of Ref.~\cite{DzubaPNCsd2001} were found to be much better than originally claimed, particularly for transitions involving $d$-states (e.g., see Figs.~\ref{fig:Cs5d},
\ref{fig:ba}, \ref{fig:ra+p32}).

\begin{table}
\caption{Summary of E1 reduced matrix elements ($ea_0$) extracted from experiment in this work by combining measurements of lifetimes ($\tau$), branching fractions, and other E1 matrix elements with our theoretical ratios (see text for details). 
Except where noted, the uncertainty is dominated by experiment.\label{tab:extracted}}
\begin{ruledtabular}
\begin{tabular}{lll}
\multicolumn{1}{l}{Transition} &  \multicolumn{1}{l}{$|\bra{a}|d|\ket{b}|$}  & Method\\
\hline
\multicolumn{3}{c}{Ca\+}\\
${3d_{3/2}}$--${4p_{1/2}}$ & 2.447(4) & E1~\cite{Hettrich2015,Ramm2013} + Ratio \\
\multicolumn{3}{c}{Sr\+}\\
$5s_{1/2}$--$5p_{1/2}$ & 3.076(24) & $\tau$~\cite{Pinnington1995} + Branching~\cite{Zhang2016} \\
 $5s_{1/2}$--$5p_{3/2}$ & 4.360(23) & $\tau$~\cite{Pinnington1995} + Branching~\cite{Zhang2016} \\
 $4d_{3/2}$--$5p_{1/2}$ & 3.093(15) & $\tau$~\cite{Pinnington1995} + Branching~\cite{Zhang2016} \\
 $4d_{3/2}$--$5p_{3/2}$ & 1.378(34) & $\tau$~\cite{Pinnington1995} + Branching~\cite{Zhang2016} \\
 $4d_{5/2}$--$5p_{3/2}$ & 4.175(22) & $\tau$~\cite{Pinnington1995} + Branching~\cite{Zhang2016} \\
\multicolumn{3}{c}{Cs}\\
 ${5d_{5/2}}$--${5p_{3/2}}$ & 9.650(18) & $\tau$~\cite{Pucher2020} \\
 ${5d_{3/2}}$--${6p_{1/2}}$ & $7.06(1)_{\rm ex}(4)_{\rm th}$ & $\tau$~\cite{Pucher2020} + Ratio \\
 ${5d_{3/2}}$--${6p_{3/2}}$ & $3.182(6)_{\rm ex}(17)_{\rm th}$ & $\tau$~\cite{Pucher2020} + Ratio \\
 ${5d_{3/2}}$--${7p_{3/2}}$ & 0.795(4) & $\tau$~\cite{Pucher2020} + E1~\cite{TohBeta2019,Bennett1999} + Ratio \\
 & 0.799(5) & $\tau$~\cite{Pucher2020} + E1~\cite{Safronova1999,Bennett1999} + Ratio \\
 ${5d_{5/2}}$--${7p_{3/2}}$ & 2.481(11) & $\tau$~\cite{Pucher2020} + E1~\cite{TohBeta2019,Bennett1999} + Ratio \\
 & 2.493(15) & $\tau$~\cite{Pucher2020} + E1~\cite{Safronova1999,Bennett1999} + Ratio \\
\multicolumn{3}{c}{Fr}\\
 ${8s_{1/2}}$--${7p_{1/2}}$ & 4.234(20) & $\tau$~\cite{Gomez2005} + Ratio \\
 ${8s_{1/2}}$--${7p_{3/2}}$ & 7.460(33) & $\tau$~\cite{Gomez2005} + Ratio \\
\multicolumn{3}{c}{Ra\+}\\
 ${7s_{1/2}}$--${7p_{1/2}}$ & 3.229(9) & E1~\cite{Fan2022} + Ratio \\
 ${6d_{3/2}}$--${7p_{1/2}}$ & 3.564(27) & E1~\cite{Fan2022} + Ratio \\
\end{tabular}
\end{ruledtabular}
\end{table}

\section{Conclusion}

We performed all-orders many-body calculations of E1 transition amplitudes between low-lying $s$, $p$, and $d$ states of K, Ca\+, Rb, Sr\+, Cs, Ba\+, Fr, and Ra\+, 
using the correlation potential method.
%
%
We compared our results with available high-precision experimental data, and find
excellent agreement, with typical deviations at the level of $\sim$\,0.1\% (or $\sim$\,0.001\,$ea_0$ -- 0.01\,$ea_0$).
Over half of our calculated amplitudes lie within experimental uncertainties, demonstrating the accuracy of the method.
As well as its accuracy, a major benefit of this approach is the efficiency; the entirety of the calculations presented here were performed with just hours of computation. 
It is our plan to make our \textsc{ampsci} implementation publicly available in the near future.

Furthermore, we showed that our method for gauging the theory uncertainty is robust; the comparison between theory and experiment is better than statistically expected, indicating our uncertainties are conservative.
We also compared our results to a number of previous theoretical determinations, finding good agreement among most of the high-precision calculations.
Finally, we combined our calculations of E1 ratios with recent experimental data to extract new E1 amplitudes for 
several transitions in Ca\+, Sr\+, Cs, Fr, and Ra\+, 
as summarized in Table~\ref{tab:extracted}.
Our results have implications for the accuracy analyses of atomic structure calculations and for the interpretation of atomic parity violation measurements.

\acknowledgements
This work was supported by the Australian Research Council (ARC) through DECRA Fellowship DE210101026 and Future Fellowship FT170100452.
We thank Andrei Derevianko for helpful comments.

\begingroup
\begin{table*}
\caption{Contributions to reduced E1 matrix elements (absolute values, units:\ $ea_0$) between the lowest few states of K, Rb, and Fr, and comparison with experiment ($\dagger$ means final theory value lies within experimental uncertainty). 
\label{tab:e1-atoms}}
\begin{ruledtabular}
\begin{tabular}{lrrrrrrrllD{.}{.}{1.4}}
         \multicolumn{1}{c}{K}
        &  \multicolumn{1}{c}{RHF}
        &  \multicolumn{1}{c}{$\delta V$}
        &  \multicolumn{1}{c}{$\delta\Sigma^{(\infty)}$}
        &  \multicolumn{1}{c}{Breit}
        & \multicolumn{1}{c}{QED}
        & \multicolumn{1}{c}{SR+N}
        & \multicolumn{1}{c}{$\delta\lambda$}
        &  \multicolumn{1}{c}{Final}
        &  \multicolumn{1}{c}{Expt.}
        &  \multicolumn{1}{c}{$\Delta$(\%)}                      \\
        \hline
$4p_{1/2}$--$4s$ & 4.5546 & -0.1540 & -0.2754 & 0.0000 & 0.0007 & -0.0065 & -0.0068 & 4.1125(75) & 4.106(2)~\cite{Falke2006} & 0.16\% \\
$4p_{3/2}$--$4s$ & 6.4391 & -0.2170 & -0.3899 & 0.0001 & 0.0009 & -0.0092 & -0.0097 & 5.814(11) & 5.807(3)~\cite{Falke2006} & 0.13\% \\
$5p_{1/2}$--$4s$ & 0.3117 & -0.0690 & 0.0149 & 0.0004 & -0.0005 & 0.0072 & 0.0038 & 0.2687(45) &  & \\
$5p_{3/2}$--$4s$ & 0.4561 & -0.0974 & 0.0223 & -0.0001 & -0.0006 & 0.0101 & 0.0054 & 0.3959(63) &  & \\
$4p_{1/2}$--$5s$ & 3.9741 & 0.0239 & -0.1071 & 0.0011 & -0.0009 & -0.0142 & 0.0091 & 3.886(10) &  & \\
$4p_{3/2}$--$5s$ & 5.6580 & 0.0334 & -0.1471 & 0.0000 & -0.0012 & -0.0201 & 0.0129 & 5.536(14) &  & \\
$5p_{1/2}$--$5s$ & 9.9348 & -0.0419 & -0.3805 & -0.0002 & 0.0015 & -0.0088 & -0.0210 & 9.484(21) &  & \\
$5p_{3/2}$--$5s$ & 14.0312 & -0.0590 & -0.5409 & 0.0005 & 0.0020 & -0.0123 & -0.0299 & 13.392(30) & & \\
$4p_{1/2}$--$3d_{3/2}$ & 8.5962 & -0.1307 & -0.4963 & 0.0001 & -0.0002 & -0.0090 & -0.0016 & 7.9585(40) & 7.984(35)~\cite{Arora2007a}\tablenotemark[1]&-0.3\%^\dagger \\
$4p_{3/2}$--$3d_{3/2}$ & 3.8546 & -0.0583 & -0.2218 & -0.0003 & 0.0000 & -0.0040 & -0.0007 & 3.5693(18) & 3.580(16)~\cite{Arora2007a}\tablenotemark[1]&-0.3\%^\dagger \\
$5p_{1/2}$--$3d_{3/2}$ & 8.1984 & 0.0271 & -1.0741 & -0.0069 & -0.0001 & -0.0155 & 0.0143 & 7.143(16) & & \\
$5p_{3/2}$--$3d_{3/2}$ & 3.6547 & 0.0122 & -0.4815 & -0.0026 & -0.0001 & -0.0069 & 0.0064 & 3.1821(70) & & \\
$4p_{3/2}$--$3d_{5/2}$ & 11.5637 & -0.1750 & -0.6654 & -0.0012 & -0.0001 & -0.0121 & -0.0007 & 10.7091(50) & 10.741(47)~\cite{Arora2007a}\tablenotemark[1]&-0.3\%^\dagger \\
$5p_{3/2}$--$3d_{5/2}$ & 10.9552 & 0.0364 & -1.4413 & -0.0092 & -0.0003 & -0.0207 & 0.0321 & 9.552(33) & & \\
\hline
         \multicolumn{1}{c}{Rb}
        &  \multicolumn{1}{c}{RHF}
        &  \multicolumn{1}{c}{$\delta V$}
        &  \multicolumn{1}{c}{$\delta\Sigma^{(\infty)}$}
        &  \multicolumn{1}{c}{Breit}
        & \multicolumn{1}{c}{QED}
        & \multicolumn{1}{c}{SR+N}
        & \multicolumn{1}{c}{$\delta\lambda$}
        &  \multicolumn{1}{c}{Final}
        &  \multicolumn{1}{c}{Expt.}
        &  \multicolumn{1}{c}{$\Delta$(\%)}                      \\
\hline
$5p_{1/2}$--$5s$ & 4.8189  & -0.2130 & -0.3558 & -0.0003 & 0.0019  & -0.0067 & -0.0069 & 4.2381(78)  & 4.233(2)~\cite{Leonard2015}\tablenotemark[2]   & 0.12\% \\
$5p_{3/2}$--$5s$ & 6.8017  & -0.2965 & -0.5063 & -0.0002 & 0.0027  & -0.0097 & -0.0100 & 5.982(11)   & 5.978(4)~\cite{Leonard2015}\tablenotemark[2]   & 0.06\%^\dagger      \\
$6p_{1/2}$--$5s$ & 0.3825  & -0.0952 & 0.0206  & 0.0009  & -0.0012 & 0.0112  & 0.0045  & 0.3232(58)  & 0.3235(9)~\cite{Herold2012}  & -0.14\%^\dagger    \\%
$6p_{3/2}$--$5s$ & 0.6055  & -0.1339 & 0.0340  & 0.0001  & -0.0015 & 0.0151  & 0.0064  & 0.5256(80)  & 0.5230(8)~\cite{Herold2012}  & 0.51\% \\
$5p_{1/2}$--$6s$ & 4.2564  & 0.0275  & -0.1316 & 0.0027  & -0.0024 & -0.0203 & 0.0125  & 4.145(14)   &            &        \\
$5p_{3/2}$--$6s$ & 6.1865  & 0.0360  & -0.1639 & 0.0006  & -0.0030 & -0.0283 & 0.0177  & 6.046(20)   &            &        \\
$6p_{1/2}$--$6s$ & 10.2856 & -0.0589 & -0.4803 & -0.0011 & 0.0040  & -0.0114 & -0.0248 & 9.713(25)   &            &        \\
$6p_{3/2}$--$6s$ & 14.4576 & -0.0813 & -0.6944 & 0.0001  & 0.0056  & -0.0159 & -0.0363 & 13.635(37)  &            &        \\
$5p_{1/2}$--$4d_{3/2}$ & 9.0464  & -0.2086 & -0.8214 & -0.0023 & -0.0006 & -0.0129 & 0.0187  & 8.019(19)   & 8.051(63)~\cite{Arora2007a}\tablenotemark[1]  &  -0.40\%^\dagger    \\
$5p_{3/2}$--$4d_{3/2}$ & 4.0817  & -0.0923 & -0.3711 & -0.0017 & -0.0002 & -0.0057 & 0.0089  & 3.6196(92)  & 3.633(28)~\cite{Arora2007a}\tablenotemark[1]  &  -0.37\%^\dagger    \\
$6p_{1/2}$--$4d_{3/2}$ & 6.7251  & 0.0523  & -1.6275 & -0.0151 & -0.0011 & -0.0251 & 0.0809  & 5.190(82)   &            &        \\
$6p_{3/2}$--$4d_{3/2}$ & 2.9551  & 0.0236  & -0.7313 & -0.0058 & -0.0006 & -0.0110 & 0.0358  & 2.266(36)   &            &        \\
$5p_{3/2}$--$4d_{5/2}$ & 12.2411 & -0.2761 & -1.1031 & -0.0060 & -0.0006 & -0.0175 & 0.0273  & 10.865(28)  & 10.899(86)~\cite{Arora2007a}\tablenotemark[1] & -0.31\%^\dagger    \\
$6p_{3/2}$--$4d_{5/2}$ & 8.8290  & 0.0704  & -2.1549 & -0.0201 & -0.0016 & -0.0328 & 0.1107  & 6.80(11)    &            &        \\
\hline
         \multicolumn{1}{c}{Fr}
        &  \multicolumn{1}{c}{RHF}
        &  \multicolumn{1}{c}{$\delta V$}
        &  \multicolumn{1}{c}{$\delta\Sigma^{(\infty)}$}
        &  \multicolumn{1}{c}{Breit}
        & \multicolumn{1}{c}{QED}
        & \multicolumn{1}{c}{SR+N}
        & \multicolumn{1}{c}{$\delta\lambda$}
        &  \multicolumn{1}{c}{Final}
        &  \multicolumn{1}{c}{Expt.}
        &  \multicolumn{1}{c}{$\Delta$(\%)}                      \\
        \hline
$7p_{1/2}$--$7s$ & 5.1438 & -0.3697 & -0.4851 & -0.0012 & 0.0064 & -0.0049 & 0.0001 & 4.2895(55) & 4.277(8)~\cite{Simsarian1998} & 0.29\% \\
$7p_{3/2}$--$7s$ & 7.0904 & -0.4636 & -0.7204 & -0.0011 & 0.0105 & -0.0097 & 0.0004 & 5.9065(85) & 5.898(15)~\cite{Simsarian1998} & 0.14\%^\dagger \\%
$8p_{1/2}$--$7s$ & 0.4589 & -0.1615 & -0.0170 & 0.0028 & -0.0040 & 0.0241 & 0.0008 & 0.3041(80) & & \\
$8p_{3/2}$--$7s$ & 1.0959 & -0.2184 & 0.0068 & -0.0001 & -0.0032 & 0.0264 & -0.0004 & 0.9070(85) & & \\
$7p_{1/2}$--$8s$ & 4.5340 & 0.0300 & -0.3137 & 0.0084 & -0.0075 & -0.0385 & 0.0188 & 4.231(23) & 4.234(20)~\cite{Gomez2005}\tablenotemark[3]& -0.06\%^\dagger\\%
$7p_{3/2}$--$8s$ & 7.7431 & 0.0059 & -0.2485 & 0.0011 & -0.0090 & -0.0488 & 0.0112 & 7.455(19) & 7.460(33)~\cite{Gomez2005}\tablenotemark[3] & -0.07\%^\dagger\\%
$8p_{1/2}$--$8s$ & 10.7837 & -0.1048 & -0.5753 & -0.0039 & 0.0130 & -0.0193 & -0.0144 & 10.079(16) & & \\
$8p_{3/2}$--$8s$ & 14.4326 & -0.1210 & -0.9487 & 0.0001 & 0.0205 & -0.0244 & -0.0212 & 13.338(24) & & \\
$7p_{1/2}$--$6d_{3/2}$ & 9.2215 & -0.4309 & -1.5856 & -0.0115 & -0.0055 & -0.0220 & 0.0375 & 7.204(39) & & \\
$7p_{3/2}$--$6d_{3/2}$ & 4.2831 & -0.1735 & -0.7938 & -0.0074 & -0.0029 & -0.0105 & 0.0191 & 3.314(20) & & \\
$8p_{1/2}$--$6d_{3/2}$ & 4.6250 & 0.1281 & -2.2134 & -0.0290 & -0.0091 & -0.0554 & 0.0673 & 2.514(71) & & \\
$8p_{3/2}$--$6d_{3/2}$ & 1.6874 & 0.0574 & -0.9619 & -0.0091 & -0.0039 & -0.0213 & 0.0280 & 0.777(29) & & \\
$7p_{3/2}$--$6d_{5/2}$ & 12.8041 & -0.5081 & -2.1149 & -0.0255 & -0.0066 & -0.0364 & 0.0670 & 10.180(69) & & \\
$8p_{3/2}$--$6d_{5/2}$ & 4.8747 & 0.1647 & -2.5077 & -0.0335 & -0.0097 & -0.0587 & 0.1058 & 2.54(11) & &  
\end{tabular}
\end{ruledtabular}
\tablenotemark[1]{Combined experiment~\cite{Miller1994} and theory~\cite{Arora2007a}};
\tablenotemark[2]{Average of Refs.~\cite{Volz1996,Simsarian1998,Gutterres2002}};
\tablenotemark[3]{Extracted in this work combining experiment~\cite{Gomez2005} with theoretical ratios}.
\end{table*}
\endgroup

\begingroup 
\begin{table*}
\caption{Contributions to reduced E1 matrix elements (absolute values, units:\ $ea_0$) between the lowest few states of Ca\+, Sr\+, Ba\+, and Ra\+, and comparison with experiment ($\dagger$ means theory value lies within experimental uncertainty). 
A negative RHF value indicates the sign changes between that and final theory value.
\label{tab:e1-ions}} 
\begin{ruledtabular}
\begin{tabular}{lrrrrrrrllD{.}{.}{1.4}}
         \multicolumn{1}{c}{Ca\+}
        &  \multicolumn{1}{c}{RHF}
        &  \multicolumn{1}{c}{$\delta V$}
        &  \multicolumn{1}{c}{$\delta\Sigma^{(\infty)}$}
        &  \multicolumn{1}{c}{Breit}
        & \multicolumn{1}{c}{QED}
        & \multicolumn{1}{c}{SR+N}
        & \multicolumn{1}{c}{$\delta\lambda$}
        &  \multicolumn{1}{c}{Final}
        &  \multicolumn{1}{c}{Expt.}
        &  \multicolumn{1}{c}{$\Delta$(\%)}                      \\
        \hline
$4p_{1/2}$--$4s$	&	3.2012	&	-0.1856	&	-0.1160	&	0.0001	&	0.0004	&	-0.0045	&	-0.0014	&	2.8943(30)	&	2.8928(43)~\cite{Hettrich2015}	&	0.05\%^\dagger\\
$4p_{3/2}$--$4s$	&	4.5269	&	-0.2614	&	-0.1639	&	0.0002	&	0.0006	&	-0.0064	&	-0.0020	&	4.0938(42)	&	4.092(6)~\cite{Hettrich2015}	&	0.05\%^\dagger\\
$5p_{1/2}$--$4s$	&	0.0061	&	0.0793	&	0.0086	&	-0.0004	&	0.0004	&	-0.0103	&	-0.0005	&	0.0834(37)	&	&	\\
$5p_{3/2}$--$4s$	&	-0.0081	&	0.1122	&	0.0115	&	0.0001	&	0.0006	&	-0.0144	&	-0.0007	&	0.1012(52)	&	&	\\
$4p_{1/2}$--$5s$	&	2.1084	&	0.0328	&	-0.0554	&	0.0009	&	-0.0007	&	-0.0156	&	0.0049	&	2.0753(70)	&	&	\\
$4p_{3/2}$--$5s$	&	3.0142	&	0.0455	&	-0.0769	&	0.0002	&	-0.0009	&	-0.0220	&	0.0070	&	2.9671(98)	&	&	\\
$5p_{1/2}$--$5s$	&	6.4426	&	-0.0620	&	-0.1400	&	0.0001	&	0.0008	&	-0.0095	&	-0.0061	&	6.2258(70)	&	&	\\
$5p_{3/2}$--$5s$	&	9.1006	&	-0.0872	&	-0.1984	&	0.0004	&	0.0011	&	-0.0134	&	-0.0087	&	8.794(10)	&	&	\\
$4p_{1/2}$--$3d_{3/2}$	&	3.0825	&	-0.1482	&	-0.4730	&	-0.0022	&	-0.0003	&	-0.0247	&	0.0185	&	2.452(21)	&	2.447(4)~\cite{Hettrich2015,Ramm2013}\tablenotemark[1]& 0.24\%	\\
$4p_{3/2}$--$3d_{3/2}$	&	1.3764	&	-0.0658	&	-0.2122	&	-0.0009	&	-0.0001	&	-0.0110	&	0.0083	&	1.0947(94)	&	1.092(2)~\cite{Hettrich2015,Ramm2013}	&	0.24\%	\\
$5p_{1/2}$--$3d_{3/2}$	&	-0.0063	&	-0.0579	&	0.1769	&	0.0011	&	0.0001	&	0.0091	&	-0.0101	&	0.113(11)	&	&	\\
$5p_{3/2}$--$3d_{3/2}$	&	0.0008	&	-0.0258	&	0.0783	&	0.0004	&	0.0001	&	0.0041	&	-0.0044	&	0.0534(47)	&	&	\\
$4p_{3/2}$--$3d_{5/2}$	&	4.1348	&	-0.1967	&	-0.6343	&	-0.0039	&	-0.0004	&	-0.0332	&	0.0252	&	3.291(28)	&	3.283(6)~\cite{Hettrich2015,Ramm2013}	&	0.26\%	\\
$5p_{3/2}$--$3d_{5/2}$	&	0.0011	&	-0.0768	&	0.2345	&	0.0016	&	0.0001	&	0.0121	&	-0.0135	&	0.159(14)	&	&	\\
\hline
         \multicolumn{1}{c}{Sr\+}
        &  \multicolumn{1}{c}{RHF}
        &  \multicolumn{1}{c}{$\delta V$}
        &  \multicolumn{1}{c}{$\delta\Sigma^{(\infty)}$}
        &  \multicolumn{1}{c}{Breit}
        & \multicolumn{1}{c}{QED}
        & \multicolumn{1}{c}{SR+N}
        & \multicolumn{1}{c}{$\delta\lambda$}
        &  \multicolumn{1}{c}{Final}
        &  \multicolumn{1}{c}{Expt.}
        &  \multicolumn{1}{c}{$\Delta$(\%)}                      \\
        \hline
$5p_{1/2}$--$5s$	&	3.4848	&	-0.2573	&	-0.1533	&	0.0000	&	0.0011	&	-0.0029	&	-0.0015	&	3.0710(29)	&	3.076(24)~\cite{Pinnington1995,Zhang2016}&	0.16\%^\dagger\\%
$5p_{3/2}$--$5s$	&	4.9211	&	-0.3583	&	-0.2171	&	0.0000	&	0.0016	&	-0.0043	&	-0.0022	&	4.3409(41)	&	4.360(23)~\cite{Pinnington1995,Zhang2016}&	0.44\%^\dagger\\%
$6p_{1/2}$--$5s$	&	-0.0664	&	0.1107	&	0.0072	&	-0.0008	&	0.0011	&	-0.0155	&	-0.0013	&	0.0350(55)	&	&	\\
$6p_{3/2}$--$5s$	&	0.1606	&	-0.1567	&	-0.0069	&	0.0000	&	-0.0014	&	0.0212	&	0.0019	&	0.0188(75)	&	&	\\
$5p_{1/2}$--$6s$	&	2.3751	&	0.0361	&	-0.0652	&	0.0021	&	-0.0018	&	-0.0220	&	0.0075	&	2.3319(89)	&	&	\\
$5p_{3/2}$--$6s$	&	3.4972	&	0.0466	&	-0.0840	&	0.0007	&	-0.0022	&	-0.0308	&	0.0103	&	3.438(12)	&	&	\\
$6p_{1/2}$--$6s$	&	6.8103	&	-0.0885	&	-0.1804	&	-0.0002	&	0.0022	&	-0.0120	&	-0.0082	&	6.5232(76)	&	&	\\
$6p_{3/2}$--$6s$	&	9.5775	&	-0.1223	&	-0.2584	&	0.0005	&	0.0031	&	-0.0166	&	-0.0122	&	9.172(11)	&	&	\\
$5p_{1/2}$--$4d_{3/2}$	&	3.7292	&	-0.2384	&	-0.4012	&	-0.0032	&	-0.0007	&	-0.0226	&	0.0247	&	3.088(20)	&	3.093(15)~\cite{Pinnington1995,Zhang2016}&	0.17\%^\dagger\\%
$5p_{3/2}$--$4d_{3/2}$	&	1.6572	&	-0.1035	&	-0.1816	&	-0.0013	&	-0.0003	&	-0.0099	&	0.0113	&	1.3719(90)	&	1.378(34)~\cite{Pinnington1995,Zhang2016}&	0.45\%^\dagger\\%
$6p_{1/2}$--$4d_{3/2}$	&	0.0263	&	-0.0962	&	0.1392	&	0.0022	&	0.0003	&	0.0147	&	-0.0167	&	0.070(13)	&	&	\\
$6p_{3/2}$--$4d_{3/2}$	&	0.0284	&	-0.0422	&	0.0606	&	0.0007	&	0.0002	&	0.0064	&	-0.0071	&	0.0470(57)	&	&	\\
$5p_{3/2}$--$4d_{5/2}$	&	5.0025	&	-0.3079	&	-0.5364	&	-0.0058	&	-0.0009	&	-0.0309	&	0.0341	&	4.155(27)	&	4.175(22)~\cite{Pinnington1995,Zhang2016}&	0.49\%^\dagger\\%
$6p_{3/2}$--$4d_{5/2}$	&	0.0758	&	-0.1242	&	0.1798	&	0.0029	&	0.0004	&	0.0184	&	-0.0219	&	0.131(18)	&	&	\\
\hline
         \multicolumn{1}{c}{Ba\+}
        &  \multicolumn{1}{c}{RHF}
        &  \multicolumn{1}{c}{$\delta V$}
        &  \multicolumn{1}{c}{$\delta\Sigma^{(\infty)}$}
        &  \multicolumn{1}{c}{Breit}
        & \multicolumn{1}{c}{QED}
        & \multicolumn{1}{c}{SR+N}
        & \multicolumn{1}{c}{$\delta\lambda$}
        &  \multicolumn{1}{c}{Final}
        &  \multicolumn{1}{c}{Expt.}
        &  \multicolumn{1}{c}{$\Delta$(\%)}                      \\
        \hline
$6p_{1/2}$--$6s$	&	3.8909	&	-0.3635	&	-0.2078	&	-0.0002	&	0.0020	&	-0.0012	&	0.0012	&	3.3214(43)	&	3.3251(21)~\cite{Woods2010}	&	-0.11\%	\\
$6p_{3/2}$--$6s$	&	5.4776	&	-0.4958	&	-0.2951	&	-0.0003	&	0.0031	&	-0.0024	&	0.0015	&	4.6886(60)	&	4.7017(27)~\cite{Woods2010}	&	-0.28\%	\\
$7p_{1/2}$--$6s$	&	-0.0654	&	0.1549	&	0.0150	&	-0.0014	&	0.0021	&	-0.0257	&	0.0012	&	0.0807(92)	&	&	\\
$7p_{3/2}$--$6s$	&	0.2610	&	-0.2197	&	-0.0122	&	0.0000	&	-0.0024	&	0.0341	&	-0.0010	&	0.060(12)	&	&	\\
$6p_{1/2}$--$7s$	&	2.5487	&	0.0457	&	-0.0929	&	0.0036	&	-0.0032	&	-0.0342	&	0.0098	&	2.478(14)	&	&	\\
$6p_{3/2}$--$7s$	&	3.9568	&	0.0495	&	-0.1094	&	0.0011	&	-0.0039	&	-0.0471	&	0.0130	&	3.860(19)	&	&	\\
$7p_{1/2}$--$7s$	&	7.3917	&	-0.1314	&	-0.2337	&	-0.0007	&	0.0038	&	-0.0174	&	-0.0062	&	7.0061(91)	&	&	\\
$7p_{3/2}$--$7s$	&	10.3120	&	-0.1757	&	-0.3393	&	0.0004	&	0.0057	&	-0.0237	&	-0.0101	&	9.769(14)	&	&	\\
$6p_{1/2}$--$5d_{3/2}$	&	3.7454	&	-0.3225	&	-0.3826	&	-0.0041	&	-0.0013	&	-0.0283	&	0.0295	&	3.036(31)	&	3.0413(21)~\cite{Arnold2019}	&	-0.16\%	\\
$6p_{3/2}$--$5d_{3/2}$	&	1.6354	&	-0.1333	&	-0.1751	&	-0.0016	&	-0.0006	&	-0.0117	&	0.0137	&	1.327(14)	&	1.33199(96)~\cite{Zhang2020a}	&	-0.39\%	\\
$7p_{1/2}$--$5d_{3/2}$	&	0.3513	&	-0.1395	&	0.0549	&	0.0022	&	0.0004	&	0.0185	&	-0.0159	&	0.272(17)	&	&	\\
$7p_{3/2}$--$5d_{3/2}$	&	0.1864	&	-0.0589	&	0.0224	&	0.0005	&	0.0002	&	0.0079	&	-0.0061	&	0.1523(67)	&	&	\\
$6p_{3/2}$--$5d_{5/2}$	&	5.0011	&	-0.3938	&	-0.5117	&	-0.0075	&	-0.0016	&	-0.0407	&	0.0416	&	4.087(44)	& 4.1028(25)~\cite{Zhang2020a}	&	-0.37\%\\
$7p_{3/2}$--$5d_{5/2}$	&	0.5425	&	-0.1696	&	0.0673	&	0.0024	&	0.0006	&	0.0206	&	-0.0197	&	0.444(21)	&	&	\\
\hline
         \multicolumn{1}{c}{Ra\+}
        &  \multicolumn{1}{c}{RHF}
        &  \multicolumn{1}{c}{$\delta V$}
        &  \multicolumn{1}{c}{$\delta\Sigma^{(\infty)}$}
        &  \multicolumn{1}{c}{Breit}
        & \multicolumn{1}{c}{QED}
        & \multicolumn{1}{c}{SR+N}
        & \multicolumn{1}{c}{$\delta\lambda$}
        &  \multicolumn{1}{c}{Final}
        &  \multicolumn{1}{c}{Expt.}
        &  \multicolumn{1}{c}{$\Delta$(\%)}                      \\
        \hline
$7p_{1/2}$--$7s$	&	3.8766	&	-0.4304	&	-0.2163	&	-0.0003	&	0.0038	&	0.0006	&	0.0015	&	3.2357(48)	&	3.229(9)~\cite{Fan2022}\tablenotemark[1]&	0.19\%^\dagger	\\%
$7p_{3/2}$--$7s$	&	5.3395	&	-0.5426	&	-0.3135	&	-0.0003	&	0.0065	&	0.0012	&	0.0019	&	4.4927(66)	&	4.484(13)~\cite{Fan2022}	&	0.19\%^\dagger	\\%
$8p_{1/2}$--$7s$	&	-0.1253	&	0.1811	&	0.0302	&	-0.0025	&	0.0036	&	-0.0293	&	0.0034	&	0.061(11)	&	&	\\
$8p_{3/2}$--$7s$	&	0.6251	&	-0.2563	&	-0.0228	&	-0.0003	&	-0.0032	&	0.0357	&	-0.0025	&	0.376(13)	&	&	\\
$7p_{1/2}$--$8s$	&	2.6367	&	0.0364	&	-0.1334	&	0.0064	&	-0.0053	&	-0.0376	&	0.0128	&	2.516(17)	&	&	\\
$7p_{3/2}$--$8s$	&	4.8100	&	-0.0024	&	-0.1298	&	0.0013	&	-0.0065	&	-0.0495	&	0.0140	&	4.637(20)	&	&	\\
$8p_{1/2}$--$8s$	&	7.3706	&	-0.1567	&	-0.2342	&	-0.0011	&	0.0071	&	-0.0201	&	-0.0066	&	6.959(10)	&	&	\\
$8p_{3/2}$--$8s$	&	9.8806	&	-0.1816	&	-0.3581	&	0.0012	&	0.0120	&	-0.0250	&	-0.0128	&	9.316(16)	&	&	\\
$7p_{1/2}$--$6d_{3/2}$	&	4.4462	&	-0.4648	&	-0.4407	&	-0.0057	&	-0.0030	&	-0.0210	&	0.0251	&	3.536(26)	&	3.564(27)~\cite{Fan2022}\tablenotemark[1]&-0.78\%	\\
$7p_{3/2}$--$6d_{3/2}$	&	1.8815	&	-0.1697	&	-0.2122	&	-0.0024	&	-0.0016	&	-0.0073	&	0.0128	&	1.501(13)	&	1.513(11)~\cite{Fan2022}	&	-0.78\%	\\
$8p_{1/2}$--$6d_{3/2}$	&	0.1053	&	-0.1913	&	0.1080	&	0.0052	&	0.0017	&	0.0334	&	-0.0217	&	0.041(24)	&	&	\\
$8p_{3/2}$--$6d_{3/2}$	&	0.1683	&	-0.0762	&	0.0392	&	0.0009	&	0.0006	&	0.0131	&	-0.0068	&	0.1391(80)	&	&	\\
$7p_{3/2}$--$6d_{5/2}$	&	5.8616	&	-0.4887	&	-0.5703	&	-0.0112	&	-0.0038	&	-0.0375	&	0.0389	&	4.789(42)	&	4.788(14)~\cite{Fan2022}	&	0.02\%^\dagger	\\%
$8p_{3/2}$--$6d_{5/2}$	&	0.4623	&	-0.2054	&	0.1056	&	0.0045	&	0.0017	&	0.0293	&	-0.0238	&	0.374(26)	&	&	
\end{tabular}
\end{ruledtabular}
\tablenotemark[1]{Extracted in this work via theoretical ratios; uncertainty is dominated by experiment}.
\end{table*}
\endgroup

\appendix
\section*{Appendix}

\section{Tables of E1 matrix elements}\label{sec:A-E1}

Our results
for 
K, Rb, and Fr are given in Table~\ref{tab:e1-atoms}, along with a comparison with experiment.
The results for Cs are given in Table~\ref{tab:E1-Cs} of the main text, and those for the ions Ca\+, Sr\+, Ba\+, and Ra\+ are in Table~\ref{tab:e1-ions}.

\section{All-orders Feynman technique}\label{sec:A-all-orders}

\begin{figure}
\centering
\includegraphics[width=0.18\textwidth]{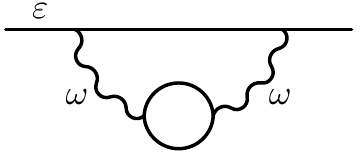}~~~~
\includegraphics[width=0.18\textwidth]{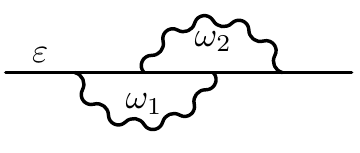}
\caption{\label{fig:Sigma2-Feyn}\small Second-order direct and exchange correlation diagrams in the Feynman technique.}
\end{figure}

The second-order Goldstone diagrams [Fig.~\ref{fig:Sigma2}] may be expressed as the two Feynman diagrams in Fig.~\ref{fig:Sigma2-Feyn} (see Ref.~\cite{Abrikosov1965}).
The internal Fermion lines represent the RHF Green's function, $ G(\en)$, external Fermion lines represent the valence wavefunction, and the photon lines represent the non-relativistic Coulomb operator, $ Q_{12}=r_{12}^{-1}$.
Evaluation of such diagrams does not require a summation over intermediate states using a basis, but rather an integration over energies. 
The loop in Fig.~\ref{fig:Sigma2-Feyn} may be described by the polarization operator: 
\begin{align}\label{eq:PolarisationOperator}
\Pi_{12}(\w) &= \int \frac{\d \en'}{2\pi}  G_{12}(\en')  G_{21}(\w+\en') ,
\end{align}
where the integral is performed analytically.
The direct part of the potential may therefore be expressed as\footnote{Subscripts are short-hand:\ $G_{12}\equiv G(\v{r}_1,\v{r}_2)$.
Integration is implied over internal indices:\ $G_{1i}G_{i2}=\int{\rm d}^3r_i G(\v{r}_1,\v{r}_i)G(\v{r}_i,\v{r}_2)$.}:
\begin{equation}\label{eq:FeynDir}
\Sigma^{(2,d)}_{12}(\en)
=
\int\frac{\d\w}{2\pi} G_{12}(\en+\w) Q_{1i}\Pi_{ij}(\w) Q_{j2},
\end{equation}
where $\en$ is the Hartree-Fock energy of the valence state.

The screening, which is enhanced by the number of electrons in the outer core shells, can be taken into account by a continued insertion of polarization loops into the Coulomb lines, as shown in Fig.~\ref{fig:Screening}.
This chain of diagrams forms a geometric series and is summed exactly:
\begin{align}\label{eq:qpq}
\widetilde Q(\w) 
&=  Q\left[1+i\,\Pi(\w) Q\right]^{-1}.
\end{align}

\begin{figure}
\centering
\includegraphics[height=0.025\textwidth]{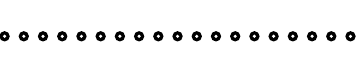}~~
\raisebox{0.015\linewidth}{\large$=$}~~
\includegraphics[height=0.025\textwidth]{img/sigma/CoulombScreening0}~~~
\raisebox{0.015\linewidth}{\large$+$}\\~~~
\includegraphics[height=0.025\textwidth]{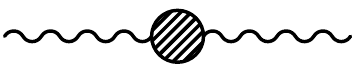}~~
\raisebox{0.015\linewidth}{\large$+$}~~
\includegraphics[height=0.025\textwidth]{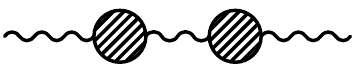}~~
\raisebox{0.015\linewidth}{\large$+$}~~
\raisebox{0.015\linewidth}{\large$\ldots$}\\
~\\
\includegraphics[height=0.03\textwidth]{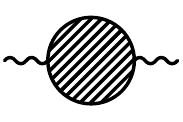}
\raisebox{0.02\linewidth}{\large$=$}
\includegraphics[height=0.03\textwidth]{img/sigma/pol0}
\raisebox{0.02\linewidth}{\large$+$}
\includegraphics[height=0.03\textwidth]{img/sigma/pol1}
\raisebox{0.02\linewidth}{\large$+$}
\includegraphics[height=0.03\textwidth]{img/sigma/pol2}
\raisebox{0.02\linewidth}{\large$+$}
\raisebox{0.015\linewidth}{\large$\ldots$} \\
\caption{\label{fig:DressedCoulomb}\small All-order screening of the Coulomb operator including the hole-particle interaction.}
\end{figure}

The hole-particle interaction [Fig.~\ref{fig:HoleParticle}] arises due to the
deviation of the RHF potential for the excited core electron in the polarization loop from that for the non-excited one~\cite{DzubaCPM1989plaEn};
the electron in the polarization loop moves in the field of $N$$-$$2$ core electrons instead of the $V^{N-1}$ potential~\cite{DzubaCPM1989plaEn}.
This is taken into account by removing the self-interaction part of the RHF potential for the excited states.
The potential that simultaneously describes the occupied core and excited states is
\begin{equation}\label{eq:V-hp}
 V = V^{N-1} - (1- P_{c})V_{\rm self} (1- P_{c}),
\end{equation}
where
$P_{c}$
is the core projection operator, and
 $V_{\rm self}$ is the self-interaction part of the RHF potential for the polarized electron.
We use this potential when forming the polarization operator.
It is typically reasonable to consider only the zero-multipolarity part of $V_{\rm self}$; the higher-multipole parts contribute negligibly.
The full screened Coulomb operator is then calculated as in Fig.~\ref{fig:DressedCoulomb}.

The direct part may then be written as 
\begin{align}
\Sigma^{(\infty,d)}_{12}
&= \int\frac{\d\w}{2\pi} G_{12}(\en+\w) Q_{1i}\Pi_{ij}(\w)\widetilde Q_{j2}(\w),
\label{eq:SigmaAll-dir}
\end{align}
which includes hole-particle interaction and screening to all orders (including dominating single, double, triple, quadruple, etc.\ excitations).
While it is possible to calculate the exchange part of the potential using the same technique, as was done in Ref.~\cite{GingesCs2002}, the double frequency integral makes this computationally demanding.
At the same time, the exchange part of the correlation potential is much smaller, and its contribution to E1 matrix elements is particularly small.
Therefore, it is reasonable to calculate this approximately, which we do via screening factors~\cite{DzubaCPM1989plaEn}.
The Coulomb operator is expanded over multipolarities 
$ Q=\sum_k  q_k$, 
and the screening factors are defined such that
$
\widetilde Q(\omega)\approx \sum_k f_k  q_k.
$
These are calculated by evaluating the direct energy correction with and without screening:
$
f_k^{(v)} = {\bra{v}\Sigma^{\rm(scr.)}_{k}\ket{v} \, / \, \bra{v}\Sigma^{}_{k}\ket{v}}.
$
We use the scaled Coulomb operators to calculate the exchange part of $\Sigma$ using the Goldstone technique as in Fig.~\ref{fig:Sigma2}.
While it is reasonable to use the same $f_k$ for each valence state, we calculate them independently for each state (i.e., including the energy dependence of $\Sigma$).

\section{Structure radiation}\label{sec:A-SR+N}

The Coulomb integrals have angular decomposition:
\begin{align}
g_{abcd} &= \int \phi^\dagger_a(\v{r}_1)\phi^\dagger_b(\v{r}_2){{r}_{12}^{-1}}\phi_c(\v{r}_1)\phi_d(\v{r}_2)\,\d^3r_1\d^3r_2\notag\\
 &= \sum_{k} A^k_{abcd} \, Q^k_{abcd}\label{eq:g-angred-1},
 \\
\widetilde g_{abcd} &\equiv g_{abcd} - g_{abdc} = \sum_{k} A^k_{abcd} \, W^k_{abcd},\label{eq:g-angred-2}
\end{align}
where the angular factor, $A$, depends on magnetic quantum numbers, while $Q$ and $W$ do not,
\begin{multline*}
A = 
(-1)^{m_a-m_b}
\sum_{q}(-1)^{k+q}{\threej{j_a}{k}{j_c}{-m_a}{-q}{m_c}\threej{j_b}{k}{j_d}{-m_b}{q}{m_d}},
\end{multline*}
and $\left(:::\right)$ is a $3j$ symbol.
Our $Q^k$ and $W^k$ differ from the $X_k$ and $Z_k$ of Ref.~\cite{Lindgren1986} by a factor $(-1)^{j_a+j_b+1}$; our definition takes advantage of the 8-fold symmetry of $Q$.

With these definitions, we can write explicitly the contribution of the structure radiation correction [Eq.~\eqref{eq:SR}]:%
{\footnotesize%
\begin{align*}%
&T^{k}_{wrvc} = \sum_{ab\mu\lambda}
(-1)^{v+r+a+b+k}
\sixj{w}{v}{k}{\lambda}{\mu}{b}
\sixj{r}{c}{k}{\lambda}{\mu}{a}
\frac{W^\mu_{wrba}Q^\lambda_{vabc}}{\en_{wr}-\en_{ba}}\\
&+\sum_{an\mu}\frac{(-1)^{w-c+k+\mu}}{[\mu]}
\sixj{w}{v}{k}{r}{c}{\mu}
\Bigg[
\frac{W^\mu_{wacn}W^\mu_{varn}}{\en_{rn}-\en_{va}}
+
\frac{W^\mu_{wnca}W^\mu_{vnra}}{\en_{wn}-\en_{ca}}
\Bigg]
\\
&+\sum_{mn\mu\lambda}
(-1)^{v+r+n+m+k}
\sixj{w}{v}{k}{\lambda}{\mu}{n}
\sixj{r}{c}{k}{\lambda}{\mu}{m}
\frac{Q^\mu_{wrnm}W^\lambda_{vcnm}}{\en_{nm}-\en_{vc}},
\\
&C^{k}_{wavc} =
\sum_{bn\mu}\frac{(-1)^{k+\mu}}{[\mu]}
\sixj{w}{v}{k}{c}{a}{\mu}
\frac{(-1)^{w-c}\,W^\mu_{wnab}W^\mu_{vncb}}{(\en_{wn}-\en_{ab})(\en_{vn}-\en_{bc})}
\\
&+\sum_{mn\mu\lambda}(-1)^{k}
\sixj{w}{v}{k}{\mu}{\lambda}{n}
\sixj{c}{a}{k}{\lambda}{\mu}{m}
\frac{(-1)^{v+a+m+n}
Q^\mu_{vmnc}W^\lambda_{wanm}}{(\en_{wa}-\en_{nm})(\en_{vc}-\en_{nm})},%
\end{align*}}%
where $\left\{:::\right\}$ is a $6j$ symbol, $\en_{ij}\equiv\en_i+\en_j$, and
$D^{k}$ is similar to $C^{k}$ with 
sums over core and excited states swapped.

\bibliography{library,note}

\end{document}